\immediate\write18{makeindex \jobname.nlo -s nomencl.ist -o \jobname.nls}

\documentclass[pdflatex]{sn-jnl}

\usepackage{blindtext, graphicx, amsmath, algorithm, algpseudocode, pifont, algcompatible, comment, layout, amsthm, amssymb}
\usepackage{enumitem}   
\usepackage{eso-pic}
\usepackage{booktabs}
\usepackage{float}
\usepackage{natbib}
\usepackage[utf8]{inputenc}
\usepackage[english]{babel}
\usepackage{hyperref} 
\usepackage{xcolor}
\usepackage{bm}
\usepackage{mathtools}
\usepackage{multirow}
\usepackage[figurename=Fig.]{caption}
\usepackage{subcaption}
\usepackage{balance}
\usepackage{supertabular}
\usepackage{booktabs}
\usepackage{makecell}

\begin{document}

\title[Blockchain-Based Resource Trading for IoT Considering Preferences]{A Blockchain-Based Distributed Computational Resource Trading Strategy for Internet of Things Considering Multiple Preferences}


\author*[1]{\fnm{Tonghe} \sur{Wang}}\email{wangth@ms.giec.ac.cn}

\author[2]{\fnm{Songpu} \sur{Ai}}\email{aisongpu@teleinfo.cn}

\author[3]{\fnm{Junwei} \sur{Cao}}\email{jcao@tsinghua.edu.cn}

\affil*[1]{\orgname{Guangzhou Institute of Energy Conversion, Chinese Academy of Sciences}, \orgaddress{\city{Guangzhou}, \postcode{510640}, \state{Guangdong}, \country{China}}}

\affil[2]{\orgname{Beijing Teleinfo Network Technology Co., Ltd.}, \orgaddress{\city{Beijing}, \postcode{101399}, \country{China}}}

\affil[3]{\orgdiv{Beijing National Research Center for Information Science and Technology}, \orgname{Tsinghua University}, \orgaddress{\city{Beijing}, \postcode{100084}, \country{China}}}


\abstract{Computational task offloading based on edge computing can deal with the performance bottleneck of traditional cloud-based systems for Internet of things (IoT).
To further optimize computing efficiency and resource allocation, collaborative offloading has been put forward to enable the offloading from edge devices to IoT terminal devices. 
However, there still lack incentive mechanisms to encourage participants to take over tasks from others.
To counter this situation, this paper proposes a distributed computational resource trading strategy addressing multiple preferences of IoT users.
Unlike most existing works, the objective of our trading strategy comprehensively considers different satisfaction degrees with task delay, energy consumption, price, and user reputation of both requesters and collaborators.
The system design uses blockchain to enhance the decentralization, security, and automation.
Compared with the trading method based on classical double auction matching mechanism, our trading strategy has more tasks offloaded and executed, and the trading results are friendlier to collaborators with good reputation.}




\maketitle

\section{Introduction}

The extensive application of sensor technologies in daily objects has given birth to the Internet of things (IoT). 
With the increasing number of IoT devices deployed,
the significant volume of the data acquired from the environment has brought great challenges to data transmission, processing, and storage services provided by centralized cloud centers. 
In many time-sensitive scenarios, e.g., healthcare, vehicular network, and smart grid, cloud centers may fail to response in a timely manner and may cause serious consequences~\cite{Wu2021Multibuffers}.
In response, the edge computing paradigm emerges as the times require.
Computation offloading is the central topic of edge computing, where IoT terminal devices with limited computational resources can transfer some or all their tasks to edge servers nearby, and edge servers can also choose to transfer their tasks further to the cloud~\cite{Vakilian2022Node}. 
In this case, local computational resources are mainly dedicated to the tasks with higher requirements for response time, therefore improving the response speed and alleviating the bottleneck of the cloud~\cite{Liyanage2021Driving}.

\subsection{Collaborative Computation Task Offloading}

To further improve the efficiency of edge computing, the concept of ``collaborative offloading\rq{}\rq{} has recently been proposed~\cite{Li2020An}.
Unlike most related works, the collaborative offloading here emphasizes the offloading between edge servers or from edge servers to nearby terminal devices with surplus computational resources. 
In this setting, users that provide resources for others are called {\em collaborators}, and users with offloading requirement are called {\em requesters}.

For collaborative offloading, it is pointed out that collaborators may still lack the motivation to take over tasks from others,
so recent works start to introduce economic incentives to encourage collaborators to contribute surplus computational resources~\cite{Ng2020Collaborative,He2020A}.
Due to different concerns of different participants, their decisions of task offloading are nevertheless determined by multiple factors in addition to economic ones.
For example, the quality of service, revenue and expense, and credibility of collaborators and requesters can simultaneously impact the experience of the participation~\cite{ai4}. 
Therefore, in order to encourage user participation and promote the practical application of the system, collaborative offloading needs to comprehensively consider the influence of multiple preferences of users.

In this paper, we model collaborative task offloading as a resource trading process between collaborators and requesters.
Its core stage, the transaction matching stage, is modeled by an optimization problem with multiple attributes (e.g., task delay, energy consumption, price, and user reputation) considered in the objectives.
By satisfying more personalized trading preferences of participants,
the matching strategy intuitively brings more incentives to its participants.

\subsection{Blockchain for Internet of Things}

IoT faces a series of security challenges, and data security, user privacy,
and service trustworthiness are the major concerns of related studies~\cite{Yousefpoor2021Secure}.
Blockchain stores data in blocks, which are shared over the system through peer-to-peer communication, verified by distributed consensus, and then connected into a chain in order.
It uses cryptographic schemes to ensure the integrity of data and messages.
Moreover, blockchain-based systems can be automated with the help of smart contracts.
Thus, blockchain can provide desirable transparency, privacy, immutability, and fault tolerance for distributed systems~\cite{Wang2022Challenges}.

Yet most blockchain-based IoT systems fail to give full play to the key features of blockchain as they only use blockchain as secure databases.
This is because the complexity of developing an entire blockchain system hinders the application of blockchain~\cite{Chan2021Simple}.
The Blockchain-as-a-Service (BaaS) design greatly reduces the implementation difficulty of blockchain-based systems and promotes the penetration of blockchain in various industries~\cite{Samaniego2016Blockchain}.
Instead of building an entire blockchain system, developers can use the ready-made blockchain interfaces and toolkits offered by the BaaS platform.
Therefore, making full use of the convenient services provided by BaaS will better fit blockchain-based collaborative offloading schemes to the decentralized nature of IoT.

In order to combine the features of blockchain more closely with IoT collaborative offloading scenario, this paper extends the BaaS application of~\cite{Nguyen2019Blockchain,ai1,ai2}.
Our work exploits many different features provided by BaaS for collaborative offloading in addition to using blockchain as a secure database.
In particular, this paper uses distributed ledger services to maintain reputation chains for participants, so as to provide incentives for the participation in collaborative transactions in a distributed manner.

\subsection{Contributions}

The main contributions of this paper are as follows:
\begin{enumerate}
\item This paper proposes a distributed computational resource trading strategy for IoT users, where the BaaS design is adopted in the architecture. Unlike most related works that simply use blockchain as a secure database, this paper takes full advantage of blockchain to promote the decentralization, reliability, and automation of resource trading.
\item This paper designs a multi-preference matching (MPM) mechanism for resource trading. The matching results between requesters and collaborators comprehensively consider the satisfaction with task delay, energy consumption, price, and reputation of each participant. 
As far as we know, few relevant studies have taken these factors into account all at once.
\item We compare our MPM mechanism with the matching strategy based on classical double auction (DA) matching mechanism~\cite{Bandara2021Flocking}. 
We perform simulation experiments to show the advantages of MPM against DA.
\end{enumerate} 

The rest of this paper is arranged as follows:
Section~\ref{sec:related} briefly reviews related works;
Section~\ref{sec:system} introduces system architecture and the workflow of blockchain-based resource trading;
Section~\ref{sec:match} explains MPM mechanism in detail;
Section~\ref{sec:numerical} conducts simulation experiments and numerical analysis by comparing MPM mechanism with DA-based matching mechanism;
Section~\ref{sec:conclusion} concludes this paper.

\section{Related Works}\label{sec:related}

To explain the motivations of our work, we provide a brief summary of the related works on blockchain-based security for IoT and incentivizing collaborative offloading in this section.

\subsection{Blockchain-Based Security for IoT}

Security issues are the major concerns of IoT-related studies, and blockchain has become popular in providing security for IoT systems.
For example, \cite{Liu2019Blockchain} combines deep reinforcement learning with blockchain to create energy-efficient data collection and secure data sharing environment.
Although it allows open participation with Ethereum public blockchain~\cite{Liu2022Blockchain}, it requires an additional Byzantine fault tolerant consensus algorithm to prevent device failure, which is not very scalable for public blockchain.
In~\cite{Chi2020A}, a blockchain layer is added to the IoT data sharing system to validate, sort, and store data trading records in a secure and reliable way. 
However, its centralized community detection subprotocol might have the possibility of privacy disclosure when calculating client similarity based on label data, which also contradicts the decentralization of blockchain.
Aiming to effectively protect user privacy, the data sharing architecture of~\cite{Lu2020Blockchain} has a permissioned blockchain module to securely store and retrieve data and a federated learning module to share data models instead of raw data. 
In spite of the accuracy and efficiency of collaborative training, the consensus process requires frequent transmission of training models between nodes outside the federated learning process, which could impose great workload to the communication network.

These works, as well as many other related works, simply use blockchain for secure data storage, and the advantages of blockchain are not fully exploited~\cite{Wang2022Challenges}.
The reason is that system designers often face huge workload when developing a sophisticated blockchain system.
BaaS can greatly reduce the difficulty of implementing blockchain-based systems by providing various basic blockchain functions.
The BaaS design in~\cite{Nguyen2019Blockchain} is deployed into the edge computing platform to support distributed resource trading for task offloading based on smart contracts.
In~\cite{ai1} and~\cite{ai2}, BaaS is integrated to undertake energy supply-demand matching in fully decentralized electric power systems.
They both use the smart contract service to provide automation, which greatly enhance the performance of energy trading. 
In view of its great convenience, BaaS is adopted in this paper in the design of a decentralized, secure, reliable, incentivizing, and automated computational resource trading system for IoT.

\subsection{Incentivizing Collaborative Offloading}

Recently, the edge computing paradigm has been extensively applied into IoT systems~\cite{Porambage2018Survey}.
To mitigate the response latency issue in edge servers, 
some works recommend collaborative offloading to allow terminal users with additional computational resources to take over the tasks of edge servers~\cite{Yu2019Collaborative,Liu2020Cooperative}.
For example, 
the concept of hybrid offloading is proposed in \cite{Yu2019Collaborative}, which extends traditional edge computing offloading to hybrid offloading that combines both edge computing offloading and device-to-device offloading.
The offloading scheme takes social relationship into consideration and tries to reduce overall execution delay while enhancing its data caching service. 
In addition, the authors of~\cite{Liu2020Cooperative} design a cooperative offloading structure that allows unmanned aerial vehicles (UAV) to execute computation tasks from each other, where a DRL method is adopted to optimize the long-term utility of the mobile edge computing network.

However, the promotion of the above collaborative offloading in practice has encountered some obstacles, because users lack incentive mechanisms to complete the computing tasks offloaded by others~\cite{Wang2020A}.
As a result, some studies include economic measures and transform collaborative offloading into computational resource trading problem.
In~\cite{Li2020An}, the intelligence and selfishness of terminal users are considered when making trading decisions. 
The offloading strategy tries to maximize social welfare by considering the cooperation between edge devices and terminal users as resource trading.
Similarly, the social welfare maximization problem in computation offloading is also studied in~\cite{He2020A}.
The authors show that their mechanism has a near-optimal competitive ratio and is able to guarantee individual rationality, truthfulness, and computational tractability.
The problem is nevertheless that in addition to economic factors, response time, energy consumption,  reputation, and many other factors may also affect the decision of computing task allocation.
The influence of multiple factors in collaborative offloading is rarely addressed in related research.
Therefore, the comprehensive consideration of various factors in our MPM-based computational resource trading method is more in line with practical requirements.

\section{Computational Resource Trading System Architecture and Workflow}\label{sec:system}

The IoT system considered in this paper is based on the classical three-layered architecture of edge computing and further extends the application scope of BaaS described in~\cite{Nguyen2019Blockchain,ai1,ai2}.
As shown by Fig.~\ref{fig:architecture}, this architecture consists of five major components:

\begin{figure}[!t]
\centering
\includegraphics[width=0.65\textwidth]{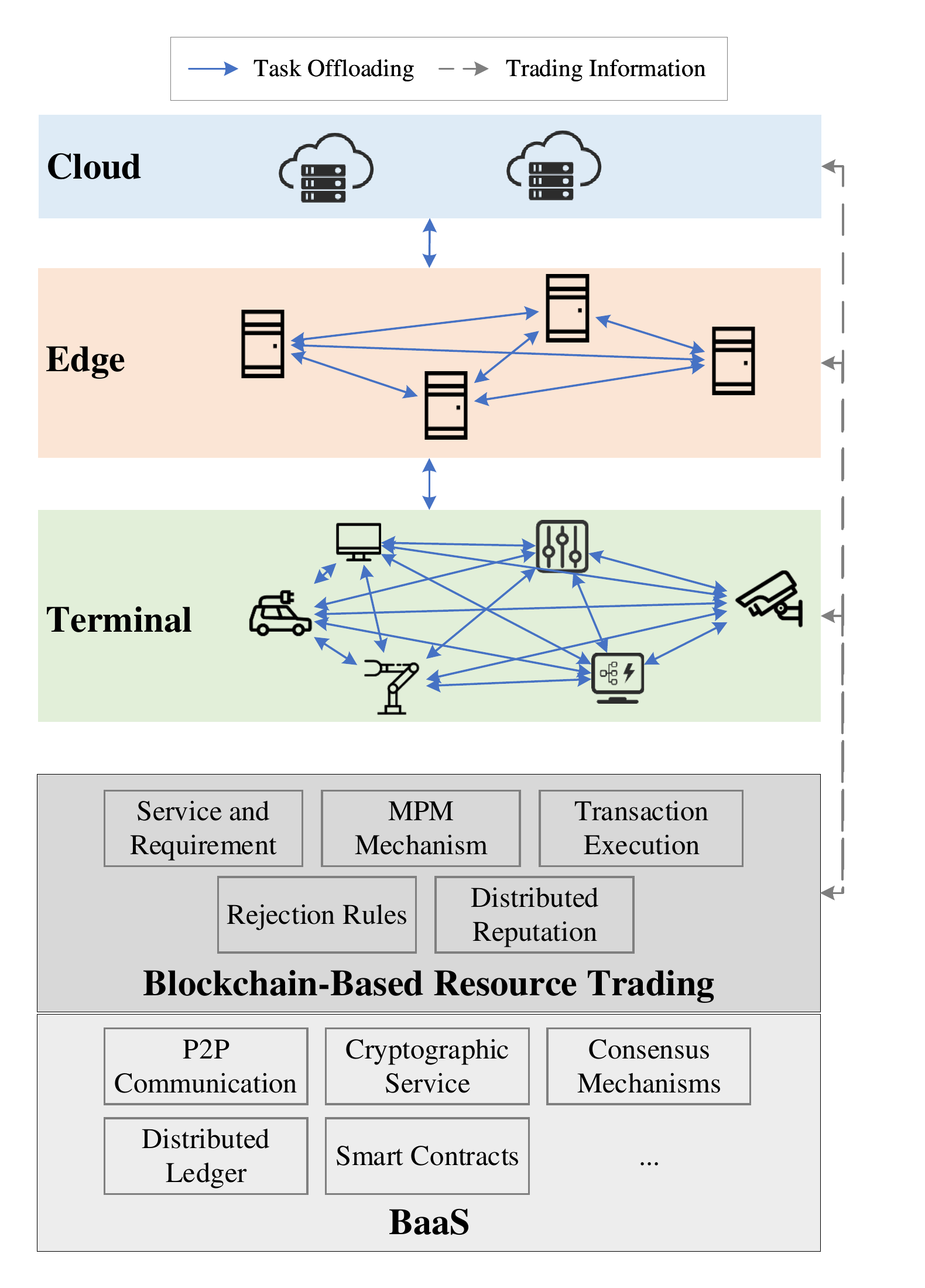}
  \vspace{-0.2cm}
\caption{System architecture.}
\label{fig:architecture}
\end{figure}

{\em Terminal.} The terminal layer is made up by IoT terminal devices embedded with sensors for data perception. 
We assume that all IoT terminals are lightweight in the sense that their computing capabilities are rather limited compared to that of edge servers.

{\em Edge.} The edge layer contains edge servers deployed near terminals. These servers have some computational resources so that they can take over the tasks offloaded from terminals. However, for economic concerns, edge servers usually have less computational resources compared to that of cloud servers. Although the transmission time between terminals and edge servers can be largely reduced, the latency caused by task queuing in edge servers cannot be neglected.

{\em Cloud.} The cloud layer has cloud servers with powerful computational capabilities, which is far away from edge servers. It is generally assumed that the cloud can process any number of tasks at the same time, but with significant transmission latency.

{\em BaaS.} BaaS platform is fundamental to achieve distributed computational resource trading for collaborative offloading. The system takes advantage of the following services of BaaS:
\begin{itemize}
\item {\bf Peer-to-peer (P2P) communication.} Messages and data are shared and retrieved transparently without the need of a central server. This makes the system more robust against single point failures. Moreover, it allows participants to join and leave the network freely.

\item {\bf Cryptographic service.} Message transmission channels and data storage are secured by various cryptographic algorithms, and user privacy is also guaranteed. 

\item {\bf Consensus mechanisms.} As the central component of blockchain, consensus is used to validate newly generated blocks that contains transaction information or reputation updates. Any change that occurs to smart contracts also needs to be validated by participants through consensus.

\item {\bf Distributed ledger.} The system uses three distributed ledgers, namely transaction chain, collaborator reputation chain, and requester reputation chain, to record transaction data and reputation scores respectively. This makes all the changes in transaction and reputation information traceable.

\item {\bf Smart contract.} The execution of resource trading and the application of reputation update rules can be automated through smart contracts, programmable scripts that automatically run when predefined conditions are met. 
\end{itemize}

\paragraph*{Blockchain-Based Resource Trading} This module is developed upon the BaaS platform. 
It contains all the functionalities that support resource trading:
\begin{itemize}
\item {\bf Service and requirements.} This submodule defines and regulates the service information submitted by collaborators and the requirement information submitted by requesters. More details are provided later in this section.

\item {\bf MPM mechanism.} Its purpose is to find matches between services of collaborators and requirements of requesters with multiple preferences of users considered. As the core of the resource trading strategy, the MPM mechanism is introduced in detail in Section~\ref{sec:match}.

\item {\bf Rejection rules.} It is possible that the matching result of one round fails to satisfy all participants. If the individual satisfaction scores of participants are not high enough, they can choose to reject the matching results. More details are given in Section~\ref{sec:reject}.

\item {\bf Transaction execution.} If  participants do not refuse the matching result, they are required to fulfill the transaction according to the matching result. The execution results will affect their reputation scores.

\item {\bf Distributed reputation.} Reputation scores are assigned to all participants to evaluate their credibility in resource trading. This submodule is simplified from the blockchain-based reputation system in~\cite{Wang2021RBT}. More details are given in Section~\ref{sec:reputation}.
\end{itemize}

In this paper, task offloading can take place between two terminals, two edge servers, and between a terminal and a server. 
During computational resource trading, collaborators can reveal the information of the resources they are willing to offer, and requesters with computational tasks can then choose the collaborators to which they will offload tasks with payments.


The distributed resource trading workflow is as follows:

{\em Step 1:} Collaborator $j$ with surplus computational resources publishes its service information including:
\begin{itemize}
\item $C_j$: size of the cache offered;
\item $f_j$: CPU frequency offered;
\item $r_j$: transmission rate offered;
\item $\epsilon_j$: maximum acceptable energy consumption per CPU cycle;
\item $op_j$: offering price, i.e., (the lowest) price of task execution offered per CPU cycle;
\item $R_j^{\textrm{c}}$: collaborator reputation score.
\end{itemize}
Meanwhile, requester $i$ with computational tasks to offload submits their offloading requirement information including: 
\begin{itemize}
\item $s_i$: size of tasks;
\item $Q_i$: CPU cycles required by tasks;
\item $\tau_i$: maximum tolerable delay of tasks;
\item $bp_i$: bidding price, i.e., (the highest) acceptable price of task execution per CPU cycle;
\item $R_i^{\textrm{r}}$: requester reputation score.
\end{itemize}

{\em Step 2:} The service of collaborator $j$ and the requirement of requester $i$ are stored in the transaction ledger.

{\em Step 3:} The smart contract of MPM, a matching mechanism considering multiple preferences, is triggered, and corresponding pre-matching results will be returned to requesters for confirmation. More details about the MPM mechanism are provided in Section~\ref{sec:match}.

{\em Step 4:} Participants can choose to reject the matching results if they are dissatisfied (see more details in Section~\ref{sec:reject}). Matching results that are not rejected are seen as accepted. Once accepted, each pre-transaction result will generate a transaction contract with the trading price calculated and will be stored into a distributed ledger in the form of smart contract provided by BaaS. Then, requesters make the payment according to their confirmed transactions. 

{\em Step 5:} On the execution time, transaction contracts will be triggered automatically, and the corresponding task offloading will take place. The reputation scores of both the requester and the collaborator will be updated according to the execution results and predefined reputation rules. Note that the submitter of an unmatched requirement can choose to either execute the tasks locally or offload the tasks further to the cloud, while the submitter of an unmatched service can choose to keep the service and wait for another round of matching.

Reputation scores of requesters and collaborators, $R_i^{\textrm{r}}$ and $R_j^{\textrm{c}}$, are defined to evaluate and regulate the behavior of requesters and collaborators. We will provide more details about our reputation system in Section~\ref{sec:reputation}.

\section{Multi-Preference Matching Mechanism}\label{sec:match}

The core of our resource trading strategy is the MPM mechanism that aims to maximizing the overall satisfaction of both requesters and collaborators considering their respective preferences.
The computational power of IoT terminal devices is much weaker than that of cloud center and edge servers. In addition, the diversity of terminal devices makes the computing tasks they need to complete and the computing services they can provide vary a lot. Task delay and offering price determine the requesters’ satisfaction with a service, while energy consumption and bidding price affect the collaborator’s experience of undertaking a requirement. At the same time, reputation can be used to evaluate the credibility of participants according to their historical behavior. 
In this case, expressing their preferences for multiple can better meet the matching requirements of both parties, so as to provide more personalized services.

The operation of our MPM mechanism also relies on the services provided by BaaS.
First, matching results are programmed into smart contracts and encapsulated into blocks. These blocks will be stored into a distributed ledger, called the {\em transaction chain}, once they are validated through distributed consensus.
Second, reputation scores for collaborators and requesters are also stored in ledgers, called {\em collaborator reputation chain} and {\em requester reputation chain} respectively. 
Participants can query each other's latest reputation scores and track the corresponding update history.
Furthermore, the rules for reputation update are also programmed into smart contracts, and any addition, deletion, and modification of these rules needs to be validated by participants through distributed consensus.

Suppose there are $m$ requesters and $n$ collaborators in a matching round.
For the sake of simplicity, we assume that no participant is both a collaborator and a requester at the same time.
Moreover, we assume that each requester submits one requirement, and each collaborator submits one service.
These assumptions are for mathematical convenience and can be easily removed by assigning unique identifiers to different roles, services, and requirements of a participant if otherwise.

We use matrix $X=(x_{ij})_{m\times n}$ to represent the result of one round of matching,
where $x_{ij}\in[0,1]$ represents the proportion of the tasks in requirement $i$ to be offloaded to collaborator $j$.
The MPM mechanism to decide $X$ works as follows:

{\em Step 1:} Fetch the information of the services of collaborators and the requirements of requesters from the transaction ledger. 
We require that bidding price $bp_i$ and offering price $op_j$ should fall in $[p_{\textrm{min}}, p_{\textrm{max}}]$, where $p_{\textrm{min}}$ and $p_{\textrm{max}}$ are the lowest and the highest prices allowed.
In practice, $p_{\textrm{min}}$ and $p_{\textrm{max}}$ are usually published in market rules and policies, and the trading service will periodically update and synchronize these values.
All the requirements and services whose prices are out of the range will be forcibly removed (similar mechanisms can also be found in~\cite{Li2019Decentralized,Yao2019Resource}).

{\em Step 2:} Calculate the service preference score (SPS) of each collaborator service for requester $i\in \{1,2,\ldots, m\}$, and calculate the requirement preference score (RPS) of each requester requirement for each collaborator $j\in \{1,2,\ldots, n\}$, with respect to different preferences. 
We will provide more details about the calculation of these two scores later on in Section~\ref{sec:sps} and~\ref{sec:rps}.

{\em Step 3:} Calculate the average requester satisfaction (ARS) for all requesters and the average collaborator satisfaction (ACS) for all collaborators.
The calculation will be introduced in Section~\ref{sec:satisfaction}.

{\em Step 4:} Model the matching as an optimization problem and find the solution. This step will be further explained in Section~\ref{sec:optimization}.


In the next, we will describe how the scores mentioned above are calculated.
Like many related works, we ignore the energy consumption and the time delay for collaborators to transfer the computation results back to requesters as the data sizes of the outcomes are usually very small in practice~\cite{Li2020An,Wang2019An}.
Moreover, to make our notation more compact, we define the following characteristic function:
\begin{equation}
\mathbb{I}[X]=
\begin{dcases*}
1 & Event $X$ is true;\\
0 & Otherwise.
\end{dcases*}
\end{equation}

\subsection{Service Preference Score Calculation}\label{sec:sps}

First, let $SPS(x_{ij})$ be the SPS of service $j$ for requester $i$.
It evaluates $i$\rq{}s comprehensive satisfaction with $j$ considering $i$\rq{}s preference in task delay, offering price, and collaborator reputation, which can be calculated by:
\begin{equation}
SPS(x_{ij})=\left(\sum_{k=1}^3 \phi_k\cdot sps_{i,j,k}\right)\cdot\mathbb{I}[x_{ij}\neq 0],
\end{equation}
where $\phi_k$ $(k=1,2,3)$ are significance factors, and $sps_{i,j,k}\in [0,1]$ $(k=1,2,3)$ will be explained in the next.
Significance factors are specified by requesters, which indicates the importance of a certain service preference to the requesters.
For example, if the requester pays more attention to the offering price, it could increase the corresponding significance factor $\phi_2$ and reduce $\phi_1$ and $\phi_3$.

\subsubsection{Task Delay}

Task delay is the main factor that influence the quality of service (QoS) of requesters, which composes transmission delay and computation delay:
\begin{equation}\label{eq:delay}
t_{ij}={x_{ij}s_i}/{r_j}+{x_{ij}Q_i}/{f_j}.
\end{equation}
Then $sps_{i,j,1}$, the task delay SPS of service $j$ for requester $i$, is calculated by:
\begin{equation}
sps_{i,j,1}=\left(1-{t_{ij}}/{\tau_i}\right)\cdot\mathbb{I}[t_{ij}\leq\tau_i].
\end{equation}
The shorter the task delay is, the better the QoS of the service is, and the higher $sps_{i,j,1}$ will be.

\subsubsection{Offering Price}

The offering price SPS of service $j$ for requester $i$, denoted by $sps_{i,j,2}$, is calculated by:
\begin{equation}
sps_{i,j,2}=e^{op_j-bp_i}\cdot \mathbb{I}[op_j\leq bp_i].
\end{equation}
The closer $op_j$ and $bp_i$ are, the more satisfied the requester will be with the matching result.

\subsubsection{Collaborator Reputation}

Collaborator reputation score $R_j^{\textrm{c}}$ reflects the credibility of service $j$ and directly serves as $sps_{i,j,3}$ in MPM mechanism:
\begin{equation}
sps_{i,j,3}=R_j^{\textrm{c}}.
\end{equation}

\subsection{Requirement Preference Score Calculation}\label{sec:rps}

Then, let $RPS(x_{ij})$ be the RPS of requirement $i$ for collaborator $j$.
It evaluates $j$\rq{}s comprehensive satisfaction with $i$ considering $j$\rq{}s preference in energy consumption, bidding price, and requester reputation, 
which can be calculated by:
\begin{equation}
RPS(x_{ij})=\left(\sum_{l=1}^3 \psi_l\cdot rps_{j,i,l}\right)\cdot\mathbb{I}[x_{ij}\neq 0],
\end{equation}
where $\psi_l$ $(l=1,2,3)$ are significance factors, and $rps_{j,i,l}$ $(l=1,2,3)$ will be explained in the next.
Like $\phi_k$, significance factors $\psi_l$ are specified by collaborators and indicates the importance of a certain requirement preference to the collaborators.

\subsubsection{Energy Consumption in Collaborator}

Compared to task delay, collaborators care more about their energy consumption when taking over the tasks from requesters.
The energy consumption in collaborator $i$ is mainly caused by receiving the data of the offloaded tasks and executing them, which can be calculated by:
\begin{equation}
E_{ji}=e_j^{\textrm{com}}{x_{ij}s_i}/{r_j}+e_j^{\textrm{exe}}{x_{ij}Q_i}/{f_j},
\end{equation}
where $e_j^{\textrm{com}}$ and $e_j^{\textrm{exe}}$ are the energy consumption of communication and task execution per second.
Considering $\epsilon_j$, the maximum energy consumption of the tasks offloaded to collaborator $j$, the RPS of requirement $i$ for collaborator $j$, denoted by $rps_{j,i,1}$, is calculated by:
\begin{equation}
rps_{j,i,1}=\left(1-{E_{ji}}/{\epsilon_j}\right)\cdot\mathbb{I}[E_{ji}\leq \epsilon_j].
\end{equation}
We can see that $rps_{j,i,1}$ decreases as $E_{ji}$ increases.

\subsubsection{Bidding Price}

The bidding price RPS of requirement $i$ for collaborator $j$, denoted by $rps_{j,i,2}$, is calculated by:
\begin{equation}
rps_{j,i,2}=e^{-{op_j}/{bp_i}}\cdot\mathbb{I}[bp_i\geq op_j].
\end{equation}
The larger $bp_i$ is, the higher the requester is willing to pay, and the more the collaborator can benefit.

\subsubsection{Requester Reputation}

Similarly, requester reputation reflects the credibility of requirement, which is directly regarded as $rps_{j,i,3}$:
\begin{equation}
rps_{j,i,3}=R_i^{\textrm{r}}.
\end{equation}

\subsection{Average Requester/Collaborator Satisfaction Score Calculation}\label{sec:satisfaction}

The requester and collaborator average satisfaction scores, denoted by $ARS(X)$ and $ACS(X)$ respectively, evaluate the overall satisfaction degree of all requesters and collaborators with matching result $X$.
The two scores are calculated by:
\begin{equation}
ARS(X)={1}/{m}\cdot\sum_{i=1}^m\sum_{j=1}^n SPS(x_{ij})x_{ij},
\end{equation}
\begin{equation}
ACS(X)={1}/{n}\cdot\sum_{j=1}^n\sum_{i=1}^m RPS(x_{ij})x_{ij}.
\end{equation}
By requiring $\sum_{k=1}^3 \phi_k=\sum_{l=1}^3 \psi_l=1$, $ARS(X)$ and $ACS(X)$ are also in the range of $[0,1]$.

\subsection{Modeling and Solving the Optimization Problem}\label{sec:optimization}

The objective of the MPM mechanism is to find the optimal $X=X^*$ that maximize $\mathcal{J}(X)$, the overall satisfaction of all participants:
\begin{align}
\max_{X\in [0,1]^{m\times n}} &\quad \mathcal{J}(X) &\label{eq:qp}\\
\textrm{s.t.} &\quad \sum_{j=1}^n x_{ij}\leq 1, \enspace 1\leq i\leq m, &\label{eq:constraint-task}\\
& \quad \sum_{i=1}^m s_ix_{ij}\leq C_j \enspace 1\leq j\leq n, &\label{eq:constraint-cache}\\
& \quad 0\leq x_{ij}\leq 1, \enspace 1\leq i\leq m,\enspace 1\leq j\leq n,&
\end{align} 
where 
\begin{equation}
\mathcal{J}(X)=w_1 ARS(X)+w_2 ACS(X),
\end{equation}
and $w_1$ and $w_2$ are the weights that indicate the significance of requesters and collaborators.
Constraint \eqref{eq:constraint-task} means that the total tasks offloaded by requester $i$ cannot exceed what is submitted in requirement $i$,
and constraint \eqref{eq:constraint-cache} means that the total size of the tasks offloaded to collaborator $j$ cannot exceed the cache size offered by service $j$.

Since $SPS(x_{ij})$ is a linear function to $x_{ij}$ because of the calculation of $t_{ij}$ by \eqref{eq:delay}, the optimization problem represented by~\eqref{eq:qp} is a quadratic programming problem.
Note that although the modeling and solving of this quadratic programming problem is centralized, it does not make our MPM strategy centralized.
The tasks of other steps, including service and requirement submission, matching rejection, transaction execution, and reputation evaluation, need the cooperation of all requesters and collaborators. 
The completion of these tasks cannot be delegated to any centralized individual in the system.
Therefore, according to the theory of ``decentralization scope and relativity''~\cite{Slepak2018The}, the MPM strategy is distributed if our scope covers all necessary steps of MPM.

\subsection{Rejection Rules and Transaction Execution}\label{sec:reject}

Since the MPM mechanism tries to maximize the overall satisfaction of all participants, it might be possible that some individuals are not willing to compromise their interests. In this case, they can choose not to accept the transaction based on their rejection rules. In more detail, users can also build smart contracts that specifies their criteria for rejection, e.g., the lowest acceptable RPS or SPS in their mind, fixed requester/collaborator ID, or requester/collaborator type (i.e., whether the request/service is from a terminal or an edge).
Note that these smart contracts also need to be validated by participants via distributed consensus.
Any matching result that fails to meet these criteria will be automatically rejected.
As a result, unmatched requirements and services will enter the next round of matching together with new ones.
Otherwise, the submitter of unmatched requirements can also choose to offload their tasks to the cloud center.

On the other hand, all matching transactions that are not rejected are seen as accepted. 
They will be saved in the transaction ledger as smart contracts and will come into effect immediately, automatically, and mandatorily.
It is worth noting that transaction execution may fail when the requester fails to pay the price, or the collaborator fails to provide the resource as promised.
Whether it is successful or not, the transaction execution results will be fed back to the reputation system, which could then affect the reputation scores of participants involved.

\subsection{Distributed Requester/Collaborator Reputation}\label{sec:reputation}

As mentioned before, requester reputation $R_i^{\textrm{r}}$ and collaborator reputation $R_j^{\textrm{c}}$ evaluate the credibility of requester $i$ and collaborator $j$ respectively.
This is a kind of distributed reputation mechanism where reputation scores rely on the mutual evaluation between requesters and collaborators, which can promote the collaborative regulation of the resource trading behavior.
In this paper, we adopt a similar idea of~\cite{Wang2021RBT} in the design of the distributed reputation mechanism.

\begin{figure}[!t]
\centering
\includegraphics[width=0.8\textwidth]{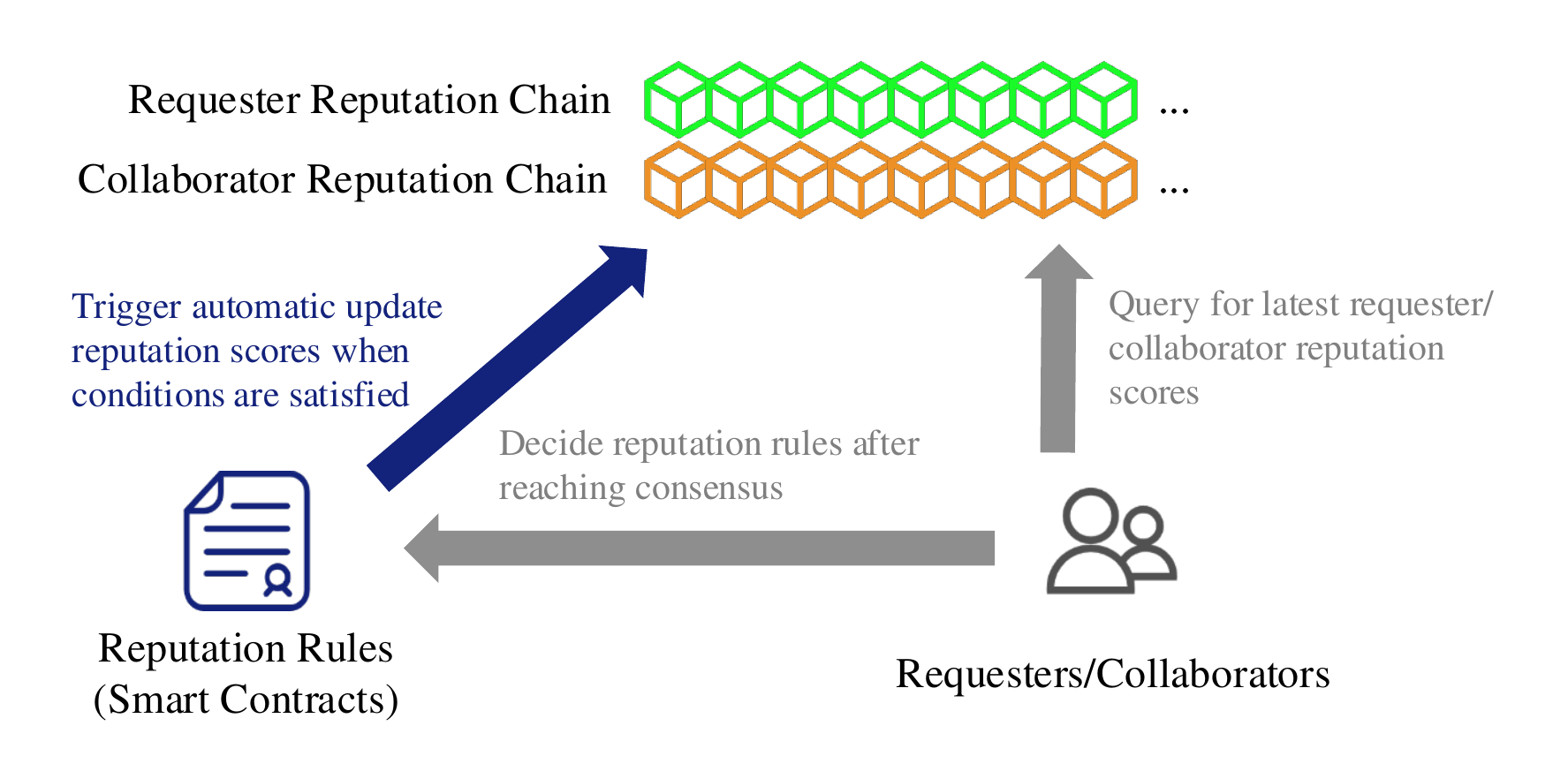}
\caption{Blockchain-based distributed reputation.}
\label{fig:reputation}
\end{figure}

The design of blockchain-based distributed reputation mechanism of this paper is shown in Fig.~\ref{fig:reputation}.
Two blockchains are maintained, one for requester reputation and the other for collaborator reputation.
These reputation scores can be queried by any participant in the system.
Reputation rules are smart contracts that specifies the methods for reputation updates and the conditions under which these updates are triggered.
All requesters and collaborators make the decision on the addition, deletion, and modification of reputation rules together by running consensus.

According to~\cite{Wang2021RBT}, blockchain can provide many favorable features for our distributed reputation mechanism.
First, reputation rules are implemented in the form of smart contracts. 
Any addition, deletion, or modification of these rules cannot come into effect until they get approved by participants by running distrbuted consensus. 
This can not only increase the transparency and traceability of reputation records, but also encourage the participation of requesters and collaborators in system regulation.
Second, these smart contracts containing reputation rules can automatically run once predefined conditions are triggered.
This circumvents the time cost and possible error of human labor in reputation update.
Furthermore, reputation records are stored in reputation chains using the linked list data structure in chronological order, and different replicas of the chains are shared among participants.
To tamper with the data in any block, the adversary also needs to modify all the data in the subsequent blocks in all replicas.
This makes it prohibitively difficult to tamper with reputation data.

\subsubsection{Reputation-Based Trading Price}

Traditional double auction method usually set the final trading price as the bidding price submitted by the buyer.
In order to enhance the fairness, we adopt the reputation-based $\alpha$-double auction~\cite{Wang2021RBT}, which calculates the trading price as follows:
\begin{equation}\label{eq:tradeprice}
tp_{ij}=\alpha\cdot bp_i+(1-\alpha)\cdot op_j,
\end{equation}
where $\alpha={R_j^{\textrm{c}}}/\left(R_i^\textrm{r}+R_j^{\textrm{c}}\right)$.
The resulting trading price will be closer to the price given by the party with the lower reputation, which is more beneficial to the party with the higher reputation.

\subsubsection{Reputation Rules}

Reputation rules are the core of the distributed reputation mechanism.
To reduce the system complexity, here we adopt the following simple rules:
\begin{itemize}
\item A new participant is assigned with an initial reputation score of $0.6$.

\item On a successful resource transaction, the reputation scores of 
both the requester and collaborator increase by $0.01$.

\item For every failed resource transaction: if requester $i$ fails to pay the trading price, then $R_i^{\textrm{r}}$ is decreased by $0.1$; if collaborator $j$ fails to provide the claimed service, then $R_j^{\textrm{c}}$ is decreased by $0.1$. 

\item The final reputation scores should be restricted in $[0,1]$.
\end{itemize}
The parameters in these rules are determined by repeated simulation tests. Every new participant is assumed to be benign and is assigned with a ``pass\rq{}\rq{} reputation score of 0.6. Moreover, the penalty of 0.1 for a failed transaction is much higher than the reward of 0.01 for a successful transaction. This can intuitively encourage participants to effectively complete each transaction.
These reputation rules are implemented as smart contracts.
Once the current trading period is over, these rules will automatically trigger reputation updates once predefined conditions are satisfied.

\section{Evaluation}\label{sec:numerical}

This section evaluates our system through simulation. In the evaluation, we mainly compare our MPM mechanism with the classical DA matching mechanism~\cite{Bandara2021Flocking}.

\subsection{Simulation Setup}

All simulation programs are written by Python 3.8 (64 bit) and are executed on a computer with {Intel\textregistered} {Core\texttrademark} i7-6500U CPU at 2.50GHz and 8GB RAM.
The optimization problem~\eqref{eq:qp} is solved via GEKKO Python library.

The ranges of parameters in the simulation are shown by Table~\ref{tab:parameter}, and all parameters are selected in their ranges uniformly at random.
The selection of these ranges is based on the works of~\cite{Nguyen2019Blockchain,Li2020An,Ng2020Collaborative,Li2021A,Wang2019An}.
We set $m=n$, and the ratio of the numbers of edge servers and IoT terminals is set by $1/30$.
In addition, by repeated adjustment and verification, we choose $w_1=w_2=0.5$, $\phi_1=\phi_3=\psi_1=\psi_3=0.36$, and $\phi_2=\psi_2=0.28$.

\begin{table}
\caption{Selection Ranges of Parameters for Simulation}
\label{tab:parameter}
\centering
\begin{tabular}{lcc}
\hline
{\bf Parameter}&{\bf Edge}& {\bf Terminal}\\
\hline
$s_i$: Size of tasks (GB) & \multicolumn{2}{c}{$[0.06, 10]$}\\
\hline
$Q_i$: CPU cycles required (Gcycle) & \multicolumn{2}{c}{$[0.6, 90]$} \\
\hline
$\tau_i$: Maximum tolerable delay (s) & $[10, 30]$ & $[5, 15]$\\
\hline

$C_j$: Cache size offered (GB) & $[5, 10]$ & $[1, 5]$\\
\hline
$f_j$: CPU frequency offered  (GHz) & $[3, 15]$ & $[1, 5]$\\
\hline
$r_j$: Transmission rate offered (Gbps) & $[0.5, 2.5]$ & $[0.1, 0.9]$\\
\hline
$\epsilon_j$: Maximum tolerable energy consumption (J) & $[150, 250]$ & $[5, 15]$\\
\hline
$e_j^{\textrm{com}}$: Unit energy consumption for transmission (J/s) & $[0.2, 0.5]$ & $[0.1, 0.3]$\\
\hline
$e_j^{\textrm{exe}}$: Unit energy consumption for execution (J/s) & $[0.75, 1.25]$ & $[0.3, 0.6]$\\
\hline

$bp_i$: Bidding price (\$/Gcycle) & \multicolumn{2}{c}{$[0.1,10]$}\\
\hline
$op_j$: Offering price (\$/Gcycle) & \multicolumn{2}{c}{$[0.1,10]$} \\
\hline

$R_i^{\textrm{c}}$, $R_j^{\textrm{r}}$: Requester/Collaborator reputation  & \multicolumn{2}{c}{$[40, 100]$}\\
\hline
\end{tabular}
\end{table}

\subsection{Double Auction}

DA is one of the most popular matching strategy in market design of various settings, e.g., computational resource allocation~\cite{Li2021Double}, energy trading~\cite{Haggi2021Multi}, data trading~\cite{Mao2020Many}, and asset market~\cite{Miklanek2020Personal}.
We believe that it is worthwhile to compare our method with DA because it is such a widely used matching mechanism.
Specifically, DA sorts the requirement list according to the ascending order of bidding prices and the service list according to the descending order of offering prices~\cite{Bandara2021Flocking}.
It then traverses each list from the top and finds a match when it encounters an offering price in the service list that is lower than the bidding price at the current location of the requirement list.
The trading price of this match will be the bidding price provided by the requester.

For each group of requirements and services generated, we execute MPM and DA respectively and compare their performance indices under the same conditions. We use solid lines to represent MPM and dashed lines to represent DA in all the figures that follows.

\subsection{Satisfaction Scores}

\begin{figure}
\centering
\includegraphics[width=0.75\textwidth]{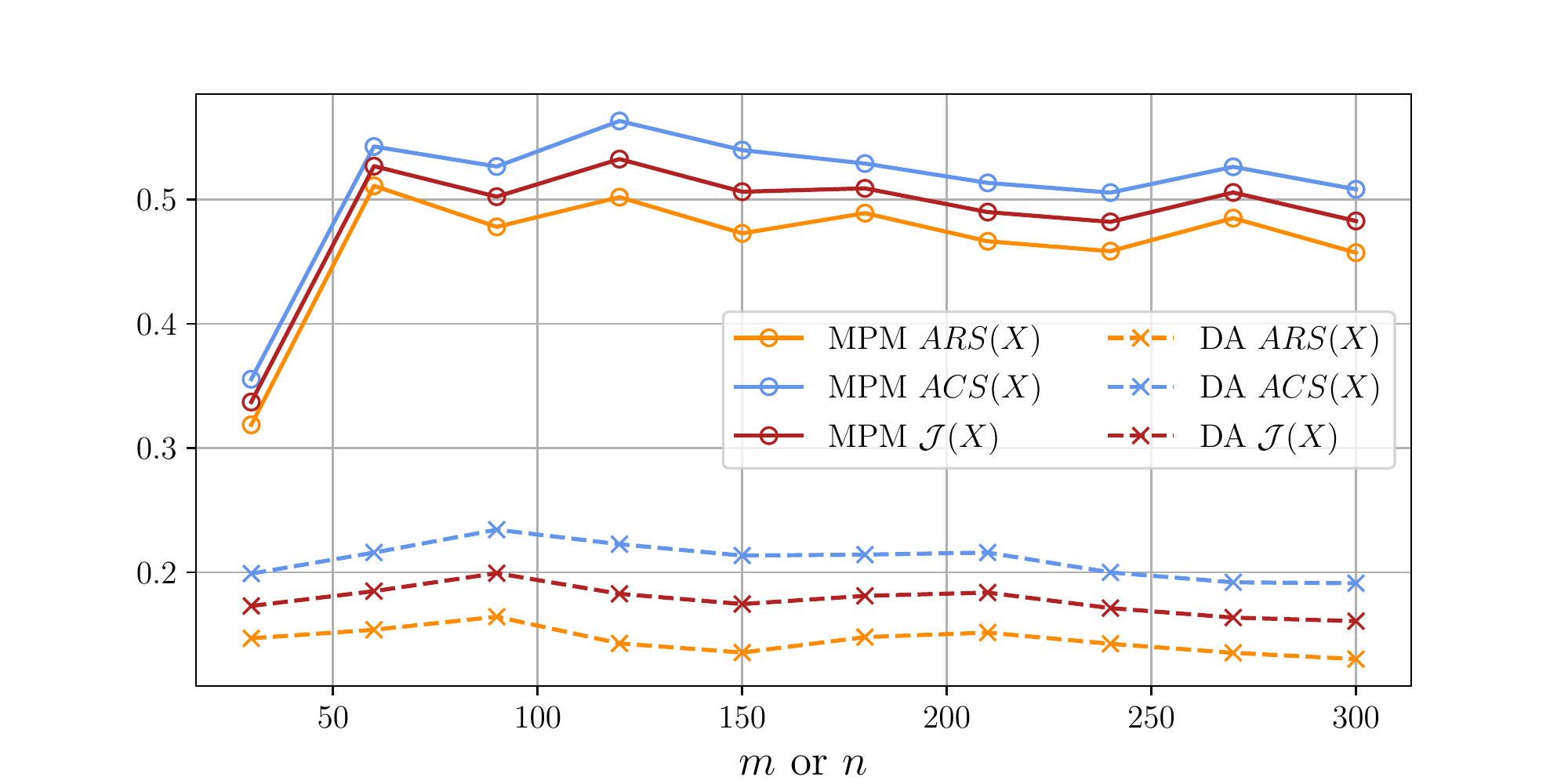}
\caption{Comparisons of ARS, ACS, and objective values.}
\label{fig:overall}
\end{figure}

Fig.~\ref{fig:overall} compares the ARS, ACS, and objective values of MPM and DA mechanisms.
We can see that there are significant gaps between the two methods in these scores:
the values of $ARS(X)$, $ACS(X)$, and $\mathcal{J}(X)$ of MPM are more than twice that of DA.
Note that these three scores evaluate the average satisfaction of requesters, collaborators, and all participants towards matching result $X$.
It can be as well interpreted that, on average, participants are more likely to accept the matching results of MPM than DA.
This intuitive provide an incentive for collaborative offloading participants when MPM mechanism is adopted.

\begin{figure}
\centering
\includegraphics[width=0.75\textwidth]{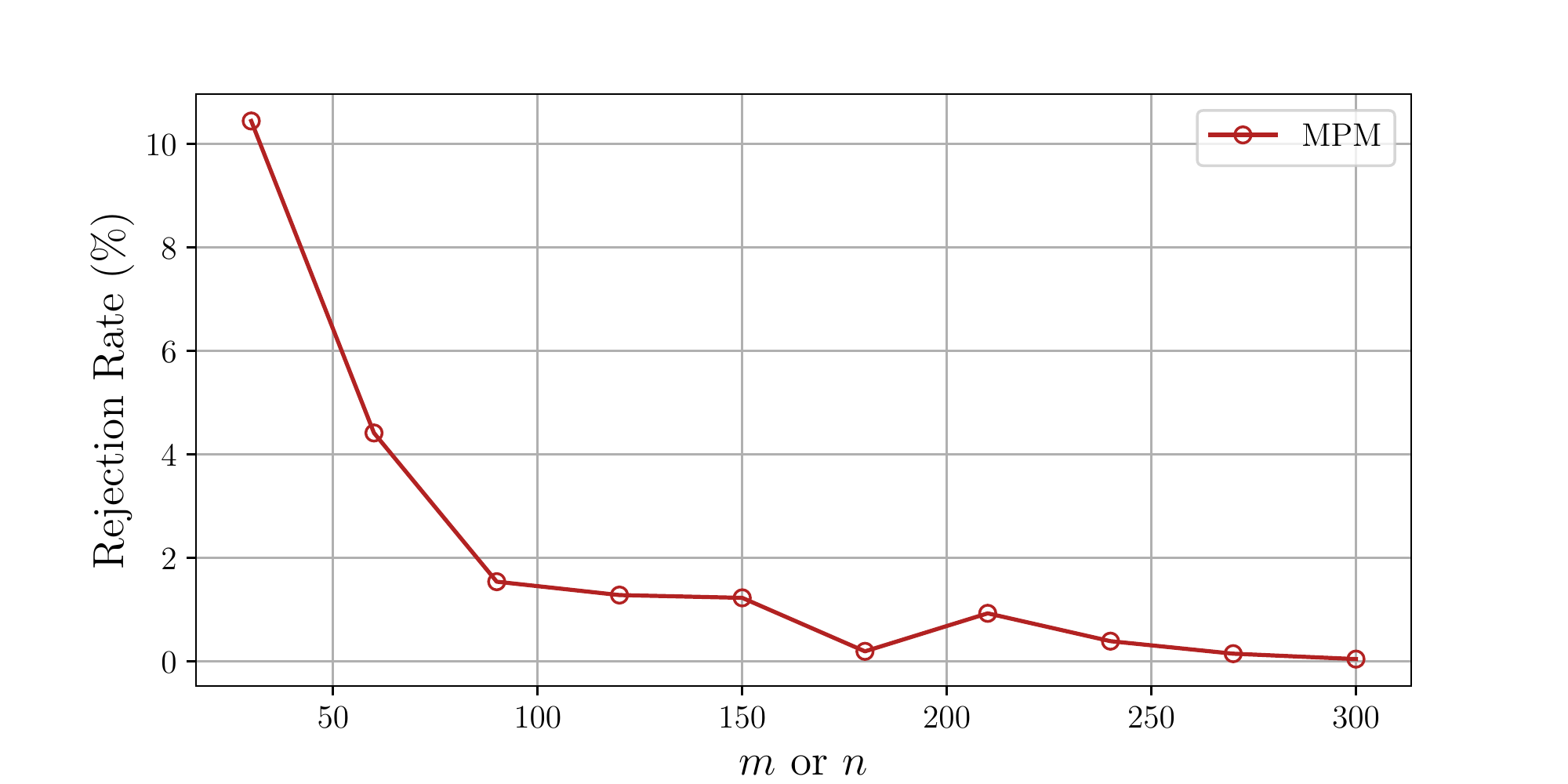}
\caption{Rejection rate of MPM.}
\label{fig:rejection}
\end{figure}

As we mentioned in Section~\ref{sec:reject}, the reasons for rejecting a matching result can varies a lot according to the individual interests of different participants.
For the sake of simplicity, we choose the lowest acceptable RPS or SPS as the rejection rule for our simulation (for other rules are rather subjective and have lower analytical value).
In more detail, we set the lowest acceptable RPS and the lowest acceptable SPS to $0.5$.
For matching result $x_{ij}>0$, if $SPS(x_{ij})<0.5$ or $RPS(x_{ij})<0.5$, then the match between requester $i$ and collaborator $j$ will be rejected, and $x_{ij}$ is set to $0$.
According to Fig.~\ref{fig:rejection}, we can find that the rejection rate of MPM decreases dramatically when the number of participants increases.
In other words, the absolutely dominant majority of the participants will be satisfied with the matching results.

\begin{figure}
\centering
\subfloat[SPS]{\includegraphics[width=0.75\textwidth]{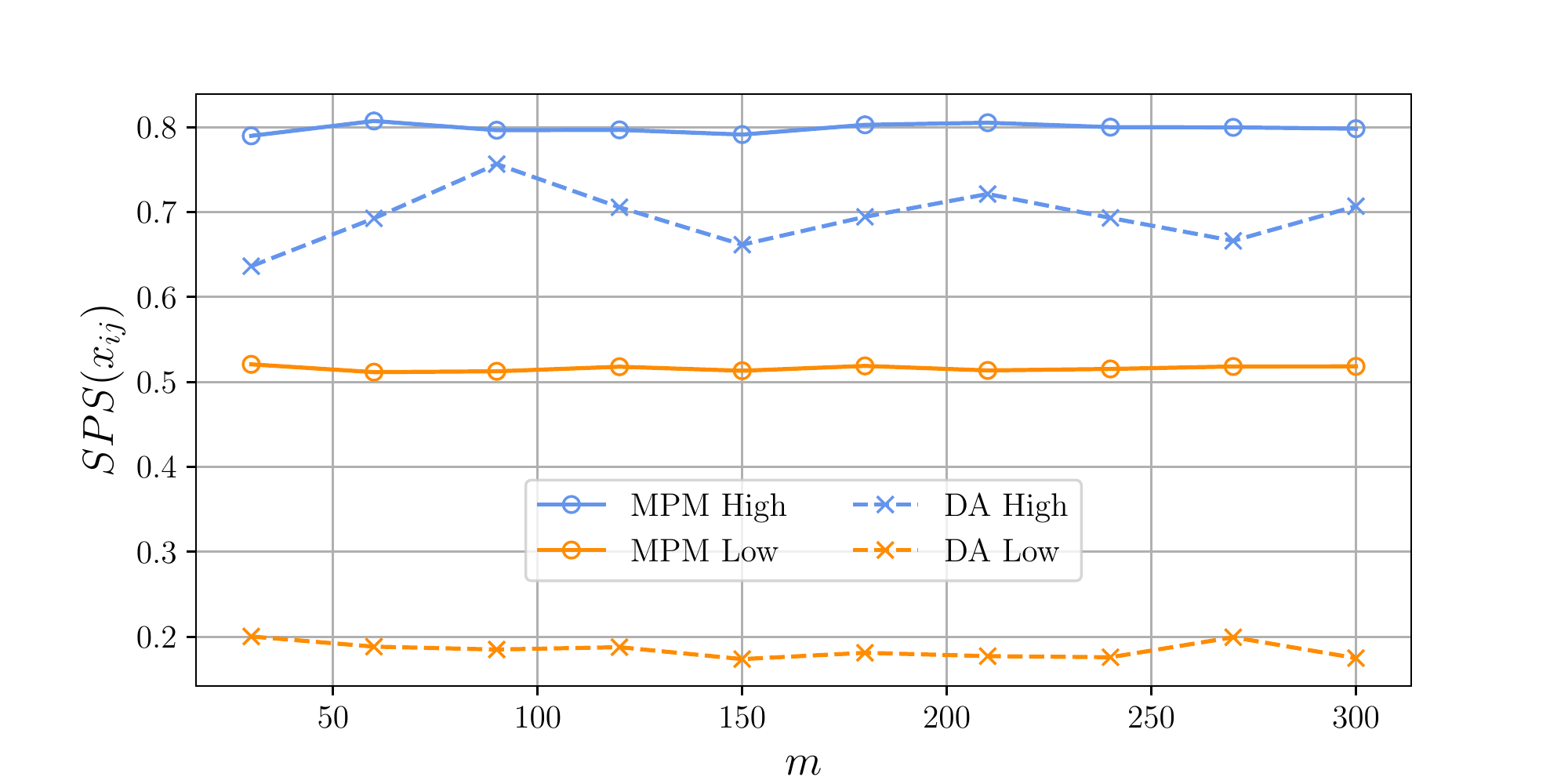}%
\label{fig:sps}}
\\
\subfloat[RPS]{\includegraphics[width=0.75\textwidth]{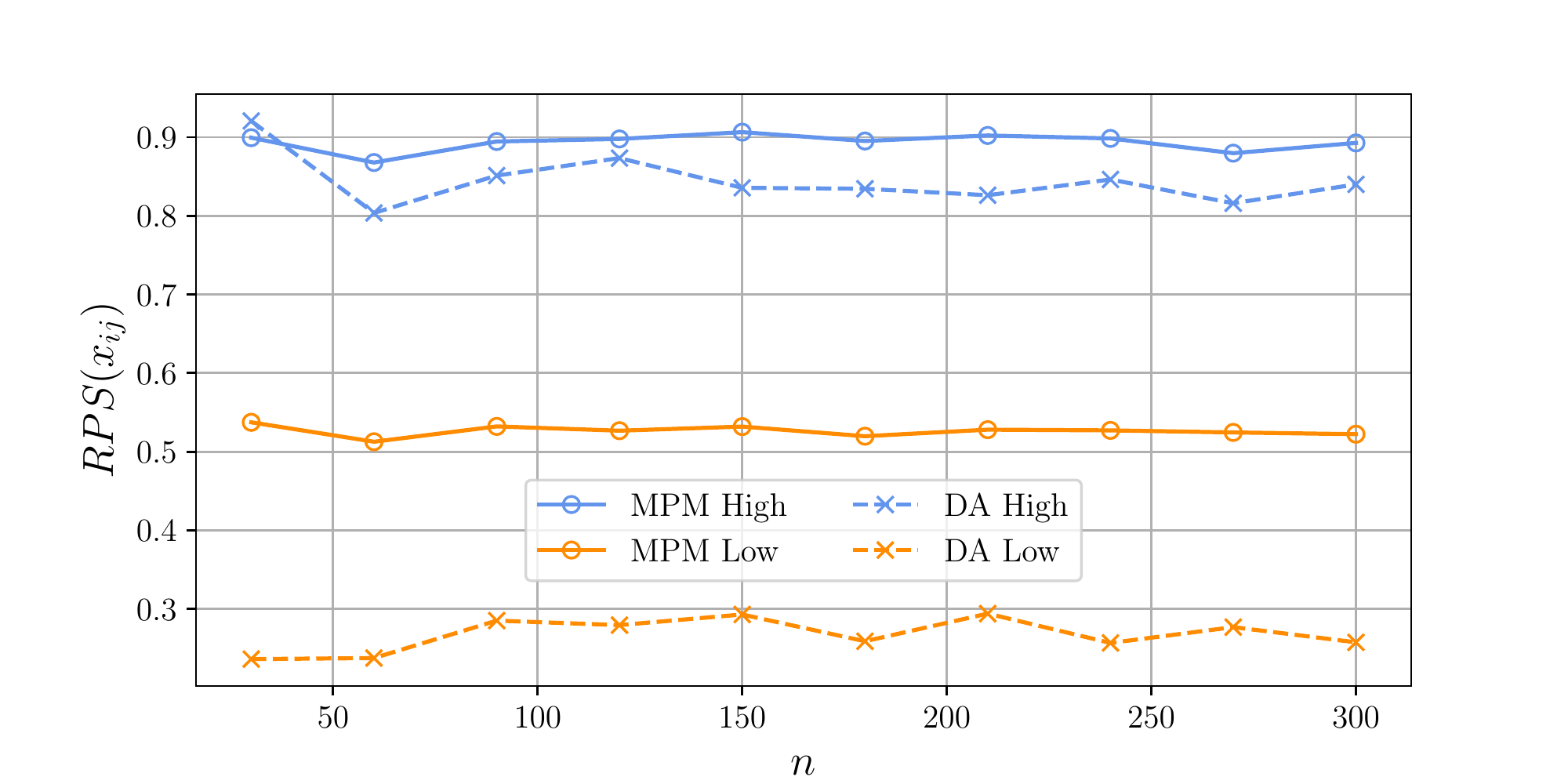}%
\label{fig:rps}}
\caption{Comparisons of high/low SPS and RPS.}
\label{fig:highlow}
\end{figure}

To further analyze the differences between the matching results of MPM and DA, we extract the average scores of SPS and RPS that rank the top $10\%$ and the bottom $10\%$ from each group of data for nonzero matching results.
We can see from the two graphs in Fig.~\ref{fig:highlow} that the high scores of SPS and RPS of MPM are slightly higher than that of DA.
Moreover, the low scores of SPS and RPS of MPM fall between $0.5$ and $0.6$, but the low scores of DA remain below $0.3$.
It indicates that more than $10\%$ matching results of DA will be rejected if participants apply the same rejection rules of MPM.

\subsection{Matching Results}

\begin{figure}
\centering
\subfloat[$X$ of MPM]{\includegraphics[width=0.42\textwidth]{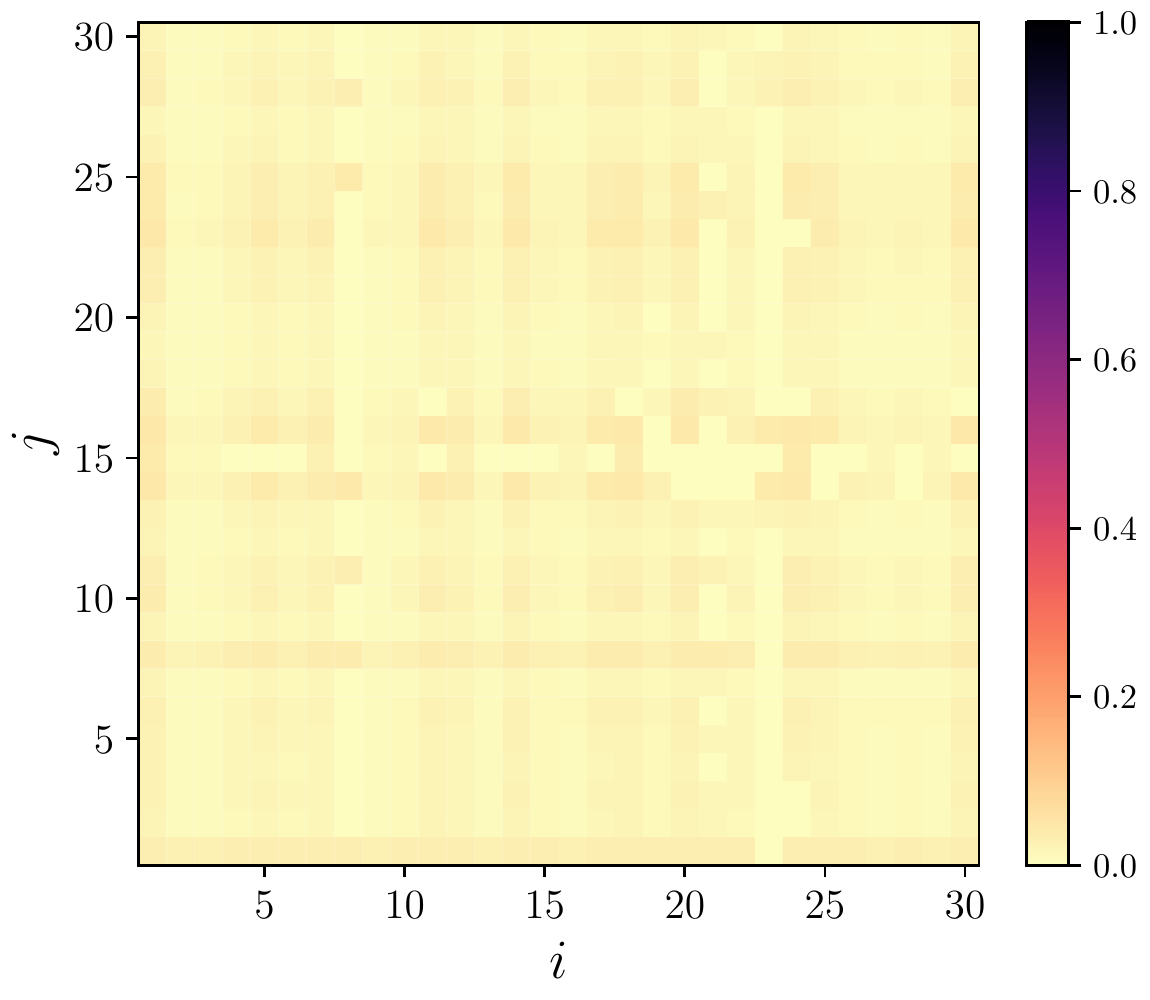}%
\label{fig:mpm_x}}
~
\subfloat[$X$ of DA]{\includegraphics[width=0.42\textwidth]{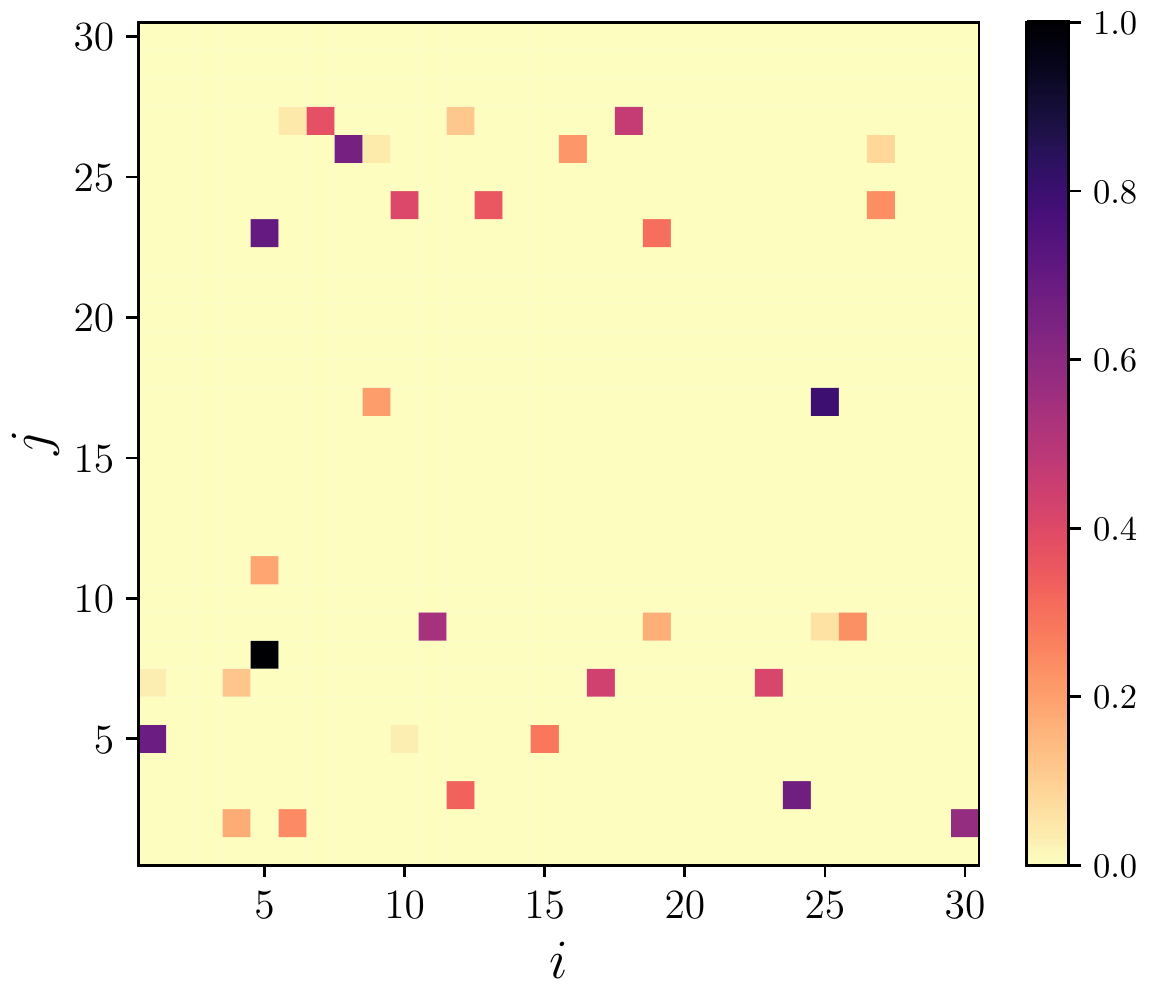}%
\label{fig:db_x}}
\caption{Comparison of matching results $X$ with $m=n=30$.}
\label{fig:x}
\end{figure}

Fig.~\ref{fig:x} visualizes an example of the matching result $X$ of both mechanisms when $m=n=30$.
In Fig.~\ref{fig:mpm_x}, the color difference of the blocks is not very obvious, but the distribution is quite uniform.
It is because that matrix $X$ of MPM does not have many zero entries, but all nonzero entries are relatively small. 
In other words, most requirements will be matched, but each matching collaborator receives a fairly small portion of these tasks.
On the other hand, Fig.~\ref{fig:db_x} has several dark-colored blocks, saying that matrix $X$ of DA has only a small number of nonzero entries.
It suggests that the number of matches DA generates is much smaller, but some of the matched collaborator may need to undertake a large proportion of the tasks.
The same pattern also holds when there are more than $30$ requesters and $30$ services.
Due to the limited space, details of these cases will be omitted here.

\subsection{Task Completion of Requesters}

\begin{figure}[!t]
\centering
\includegraphics[width=0.75\textwidth]{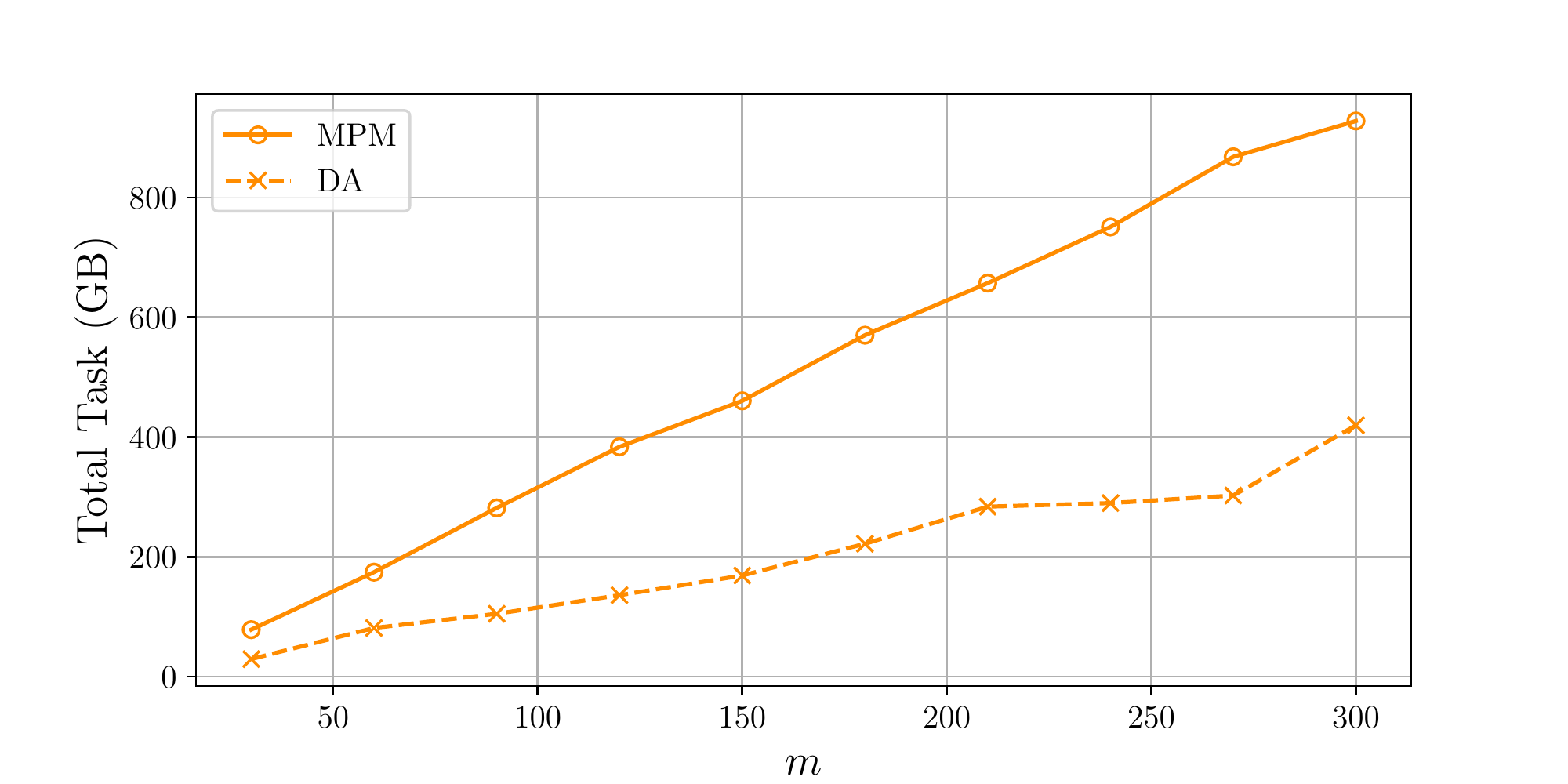}
\caption{Comparison of total sizes of tasks executed.}
\label{fig:task}
\end{figure}

\begin{figure}
\centering
\includegraphics[width=0.75\textwidth]{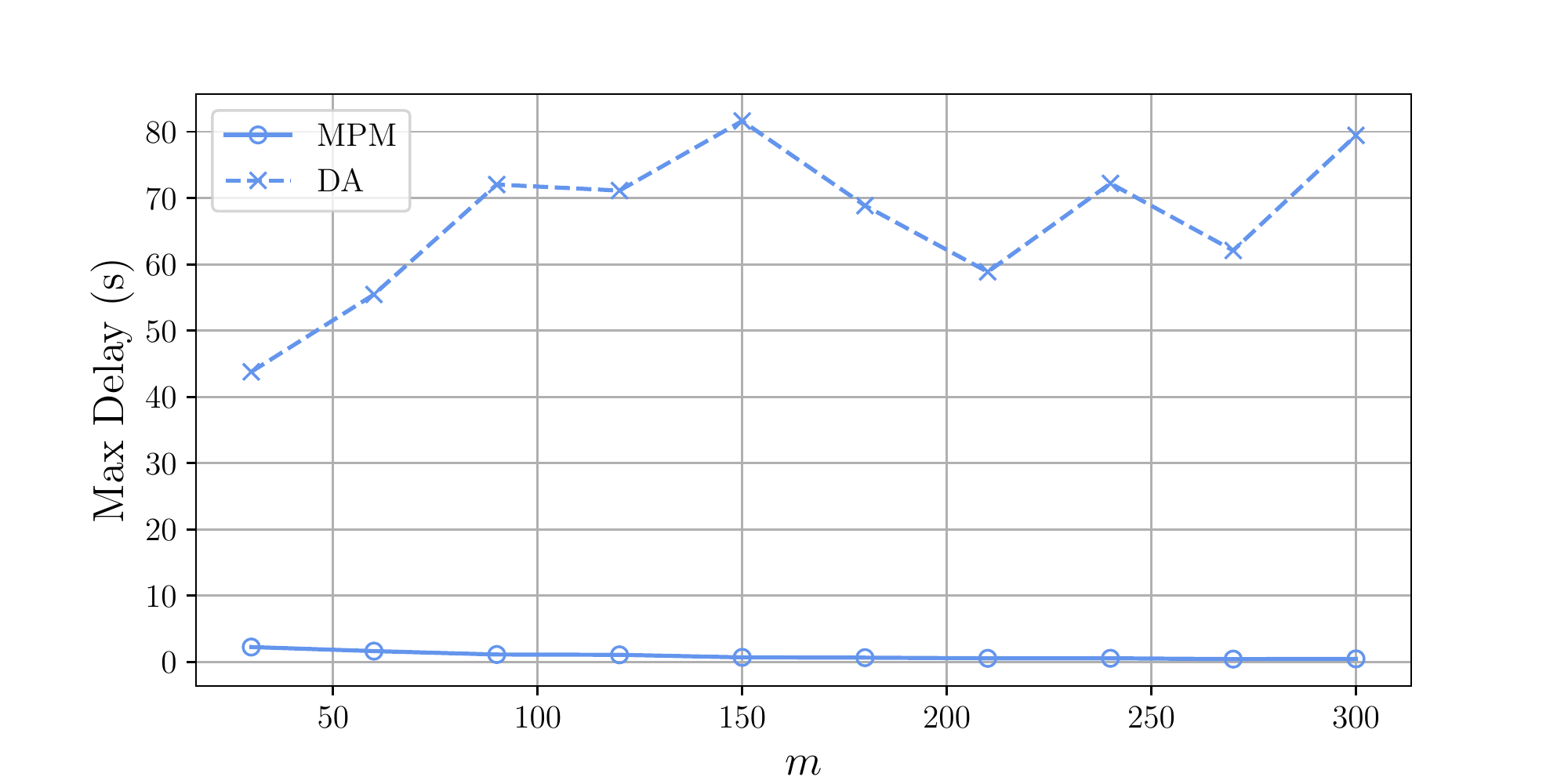}
\caption{Comparison of maximum task delays.}
\label{fig:delay}
\end{figure}

Here we observe the completion of the requesters' tasks.
Fig.~\ref{fig:task} compares the total sizes of the tasks executed of both mechanisms.
It shows that the total size of the tasks completed by using MPM is about $2.5$ times of that of DA.
On the other hand, compared with DA, Fig.~\ref{fig:delay} exhibits a reduction of more than $60$ times in the maximum task delay of MPM.
The huge gaps in Fig.~\ref{fig:task} and~\ref{fig:delay} are caused by MPM's more equal distribution of tasks among collaborators.
This is also supported by the example in Fig.~\ref{fig:x}.


\subsection{Resource Consumption of Collaborators}

\begin{figure}[!t]
\centering
\subfloat[CPU Cycles]{\includegraphics[width=0.75\textwidth]{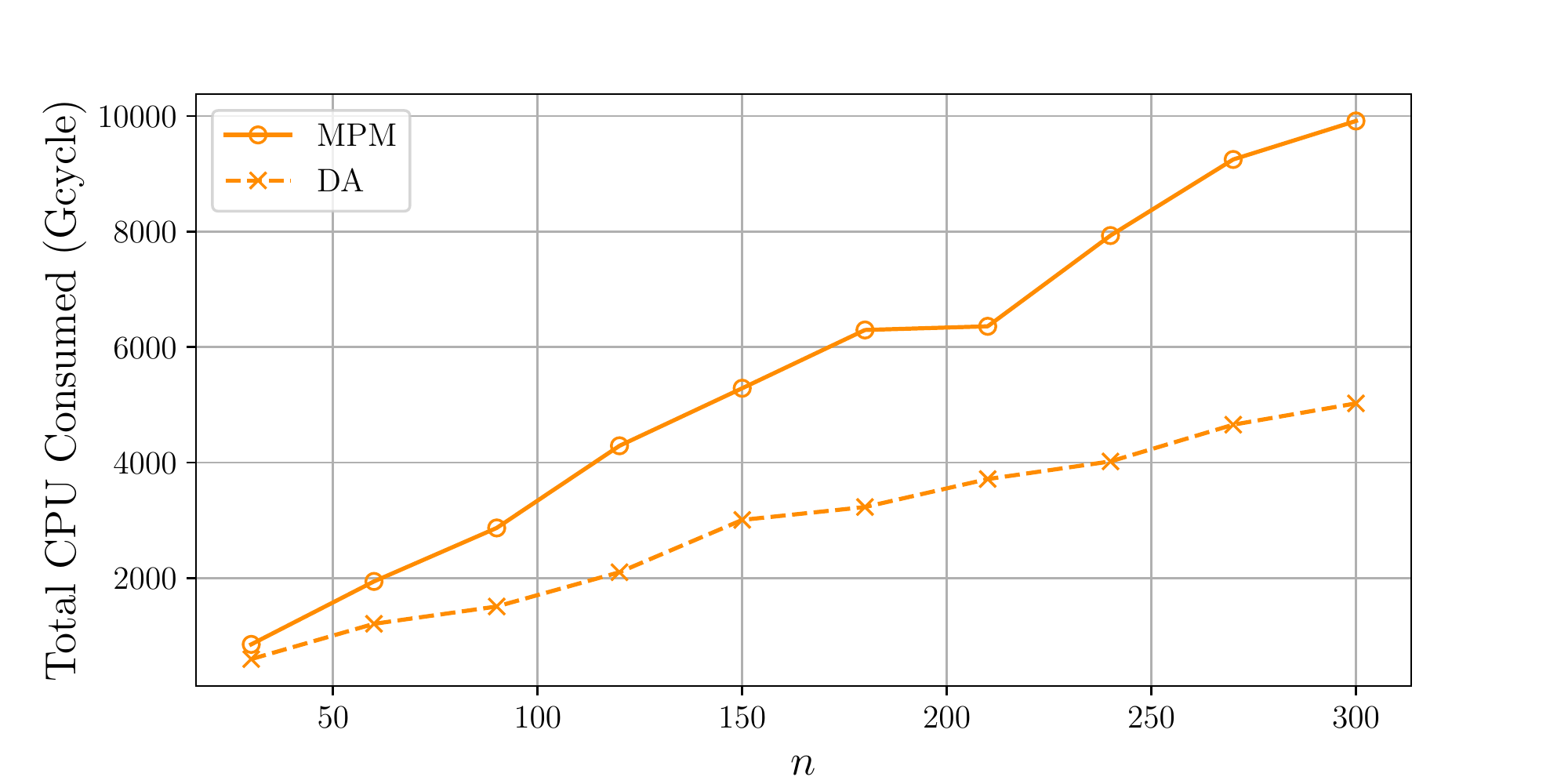}%
\label{fig:cpu}}
\\
\subfloat[Cache Sizes]{\includegraphics[width=0.75\textwidth]{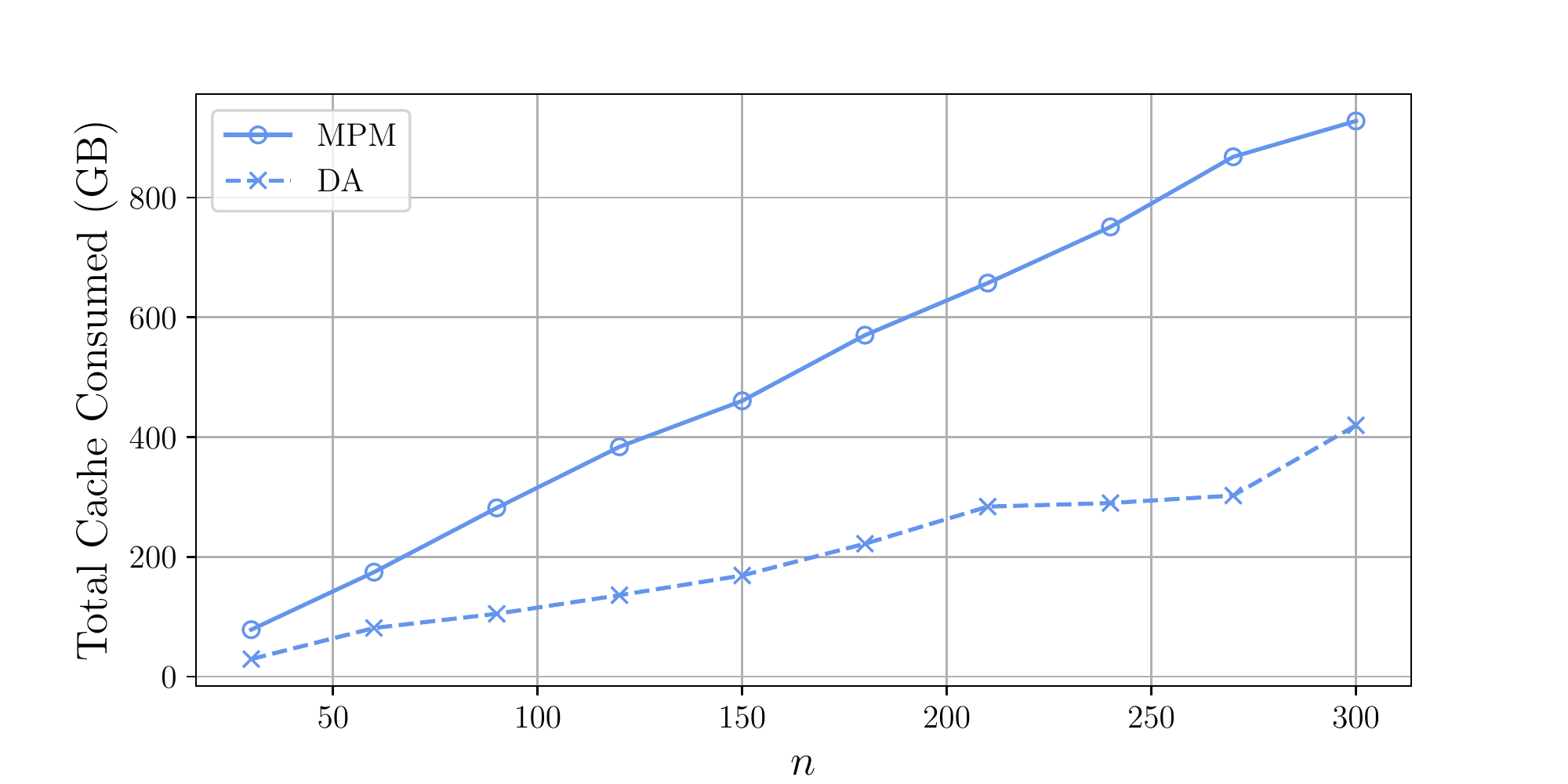}%
\label{fig:cache}}
\\
\subfloat[Energy]{\includegraphics[width=0.75\textwidth]{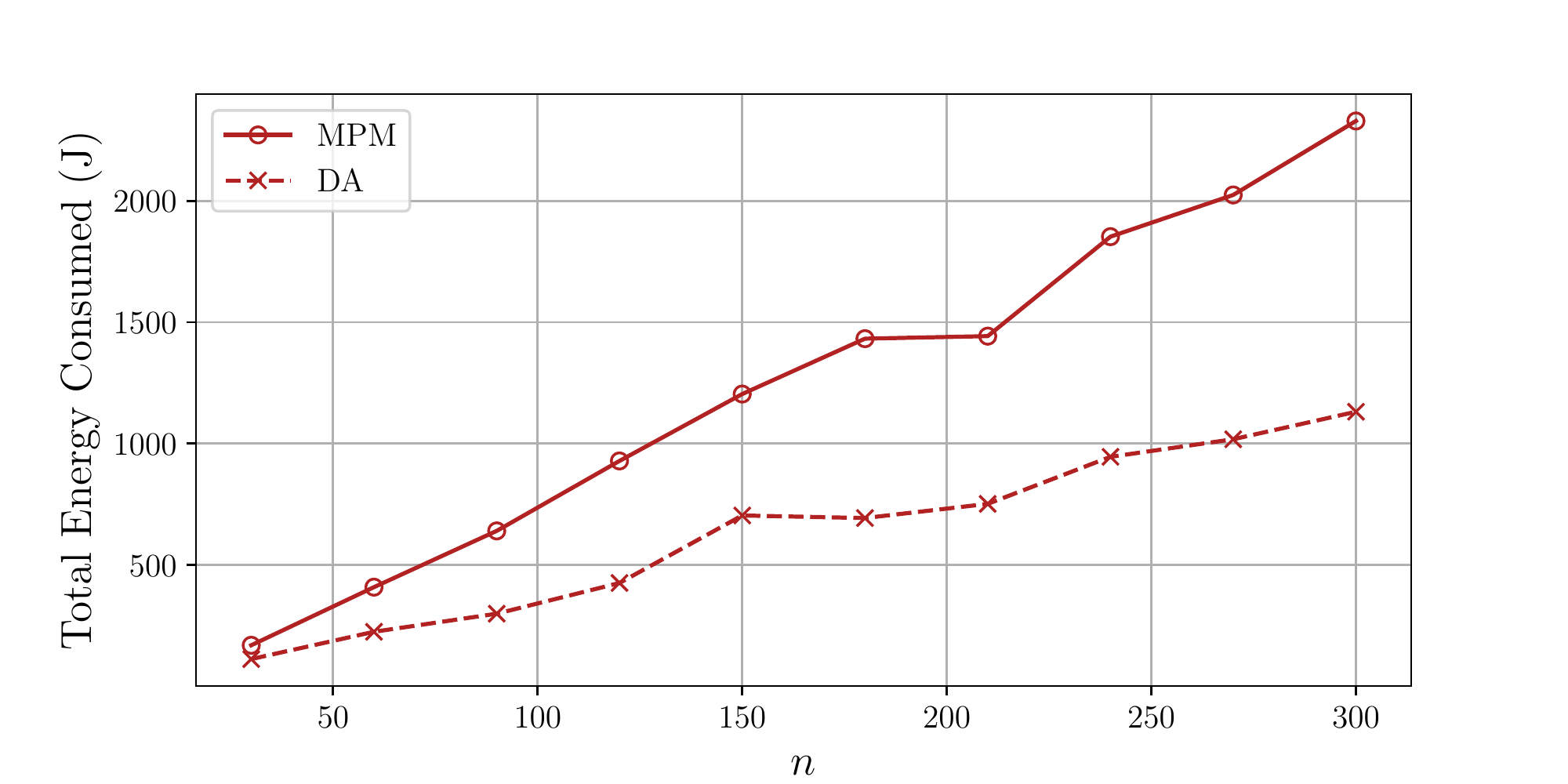}%
\label{fig:energy}}
\caption{Comparisons of resource consumption of collaborators.}
\label{fig:consumption}
\end{figure}

Next, we evaluate the resource consumption of collaborators.
Fig.~\ref{fig:consumption} compares the consumption of total CPU cycles, cache sizes, and energy of collaborators between two mechanisms.
By calculation, we find that compared with DA, MPM increases the consumption of these three resources by $89\%$, $152\%$, and $97\%$ respectively.
This significant increase in resource consumption of collaborators is because more tasks can be executed by adopting the matching results of MPM.

\subsection{Reputation and Trade Price}

\begin{figure}[!h]
\centering
\includegraphics[width=0.75\textwidth]{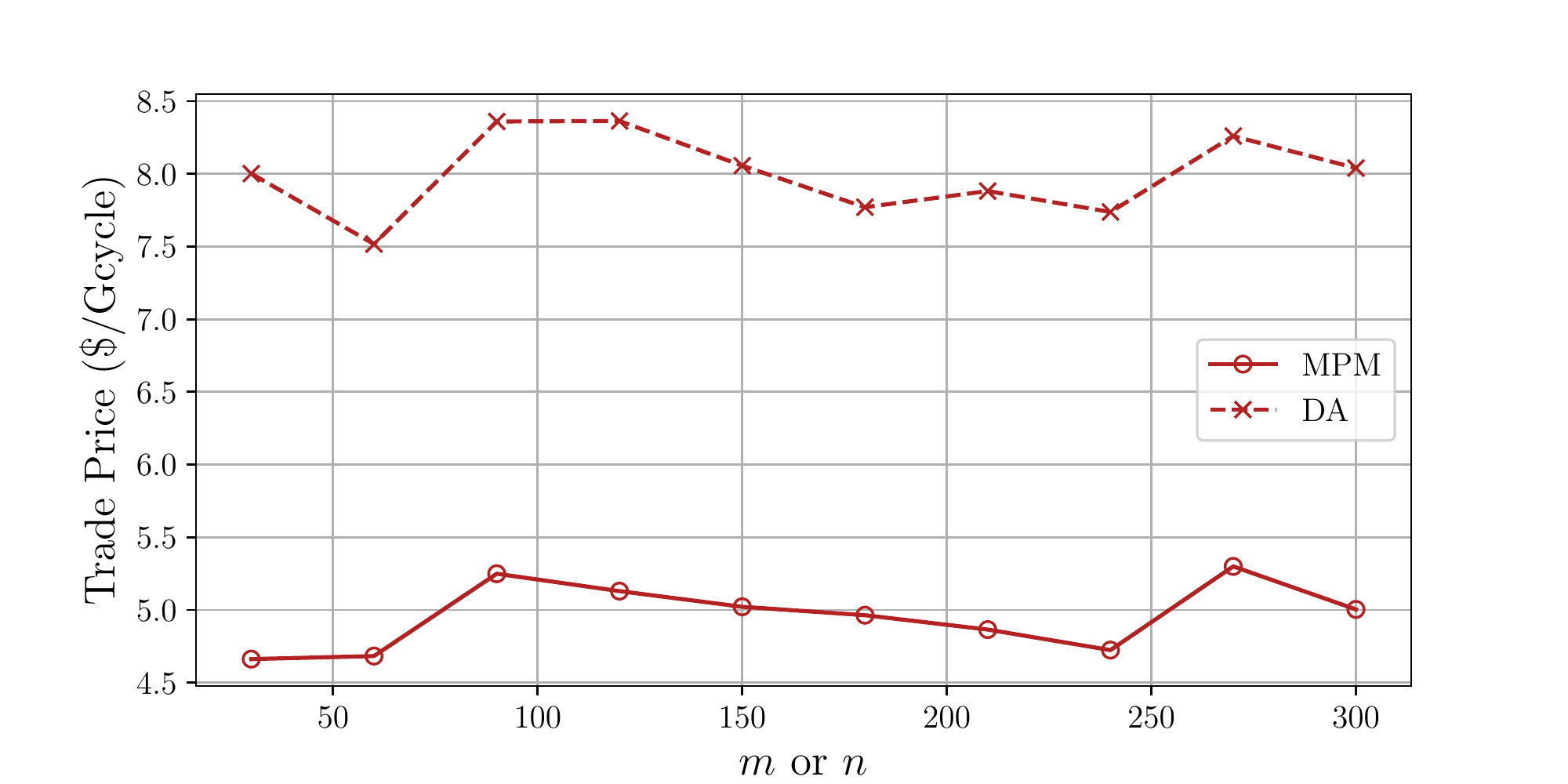}
\caption{Comparison of average trade prices.}
\label{fig:price}
\end{figure}

Fig.~\ref{fig:price} compares the average trade prices of two matching mechanisms.
The trade prices of MPM is about $37.99\%$ lower than that of DA on average.
This drop is because that MPM adopts $\alpha$-double auction in~\eqref{eq:tradeprice} where the reputation scores of both collaborator and requester are taking into account while calculating the trade price.

\begin{figure}[!t]
\centering
\subfloat[Average Cost vs. Requester Reputation]{\includegraphics[width=0.75\textwidth]{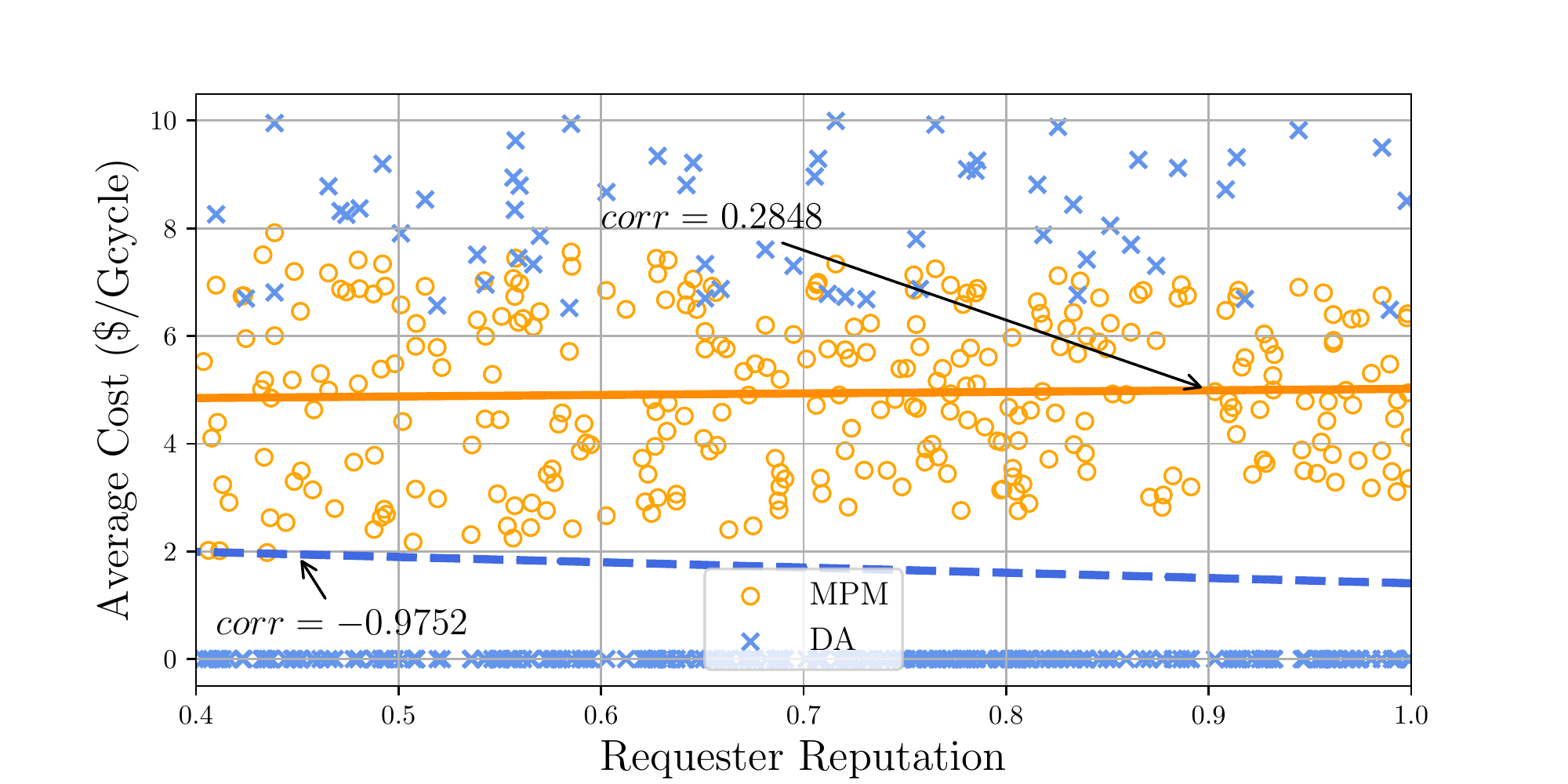}%
\label{fig:corr_r}}
\\
\subfloat[Average Income vs. Collaborator Reputation]{\includegraphics[width=0.75\textwidth]{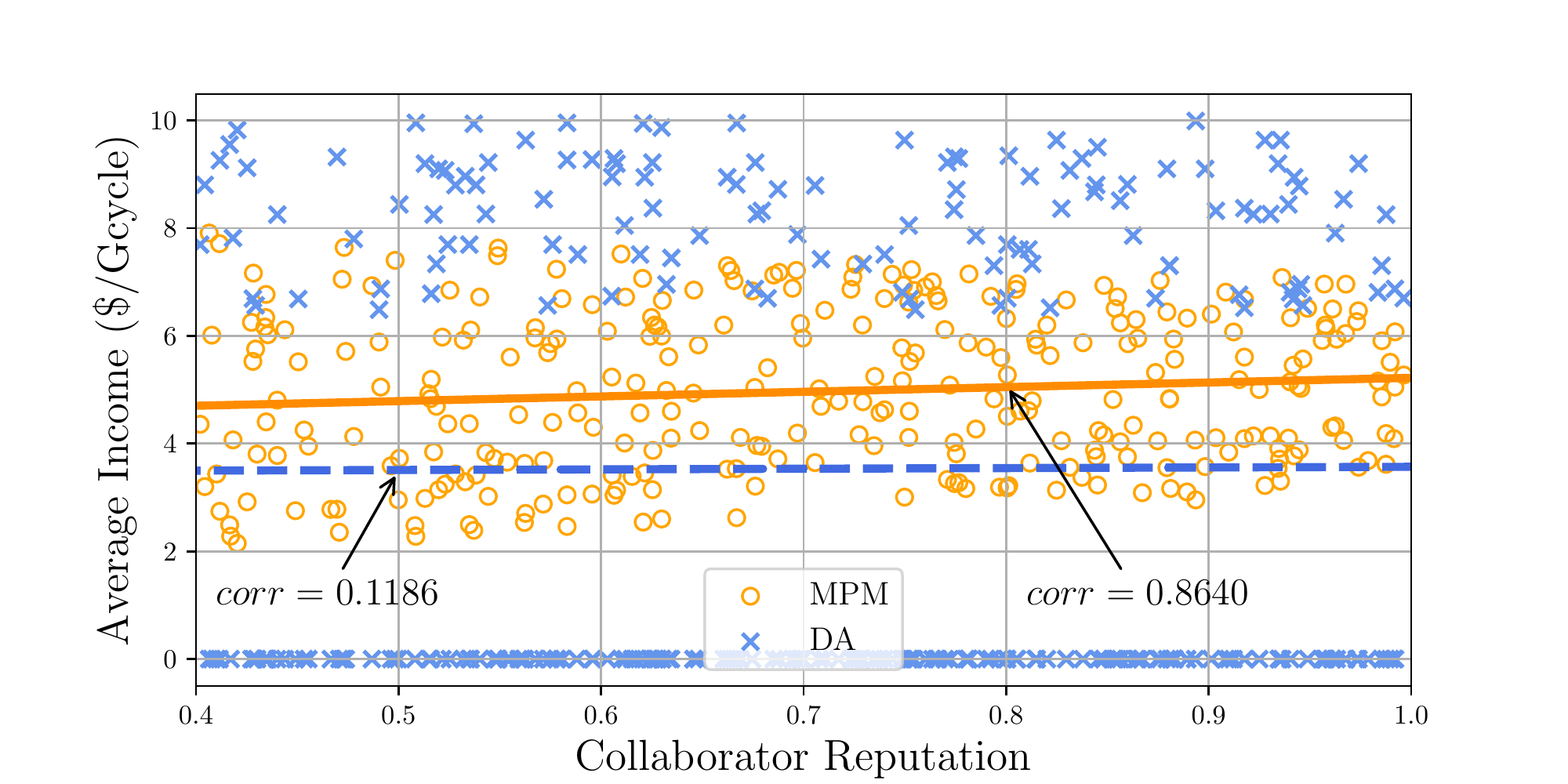}%
\label{fig:corr_c}}
\caption{Comparisons of average costs of requesters and average incomes of collaborators.}
\label{fig:corr}
\end{figure}

We then look into the relationship between requester/collaborator reputation and average cost/income.
Fig.~\ref{fig:corr} looks into the matching results of the case of $m=n=300$.
In Fig.~\ref{fig:corr_c}, we plot the scatter chart of average cost versus requester reputation and then draw the corresponding linear fitting line.
The slope of the linear fitting line, denoted by $corr$, is the correlation coefficient between average cost and requester reputation.
Since the $corr$ of DA is negative but has a greater absolute value,
it indicates that the DA mechanism is friendlier to requesters with higher reputation scores.
Similarly, Fig.~\ref{fig:corr_r} shows that the $corr$ of MPM is positive and has a slightly greater absolute value, it indicates that the MPM mechanism is more friendly to collaborators with higher reputation scores.
This also indirectly proves that our resource trading system with MPM can better incentivize the participation of collaborators through the distributed reputation system.

\subsection{Computational Efficiency}

\begin{figure}
\centering
\includegraphics[width=0.75\textwidth]{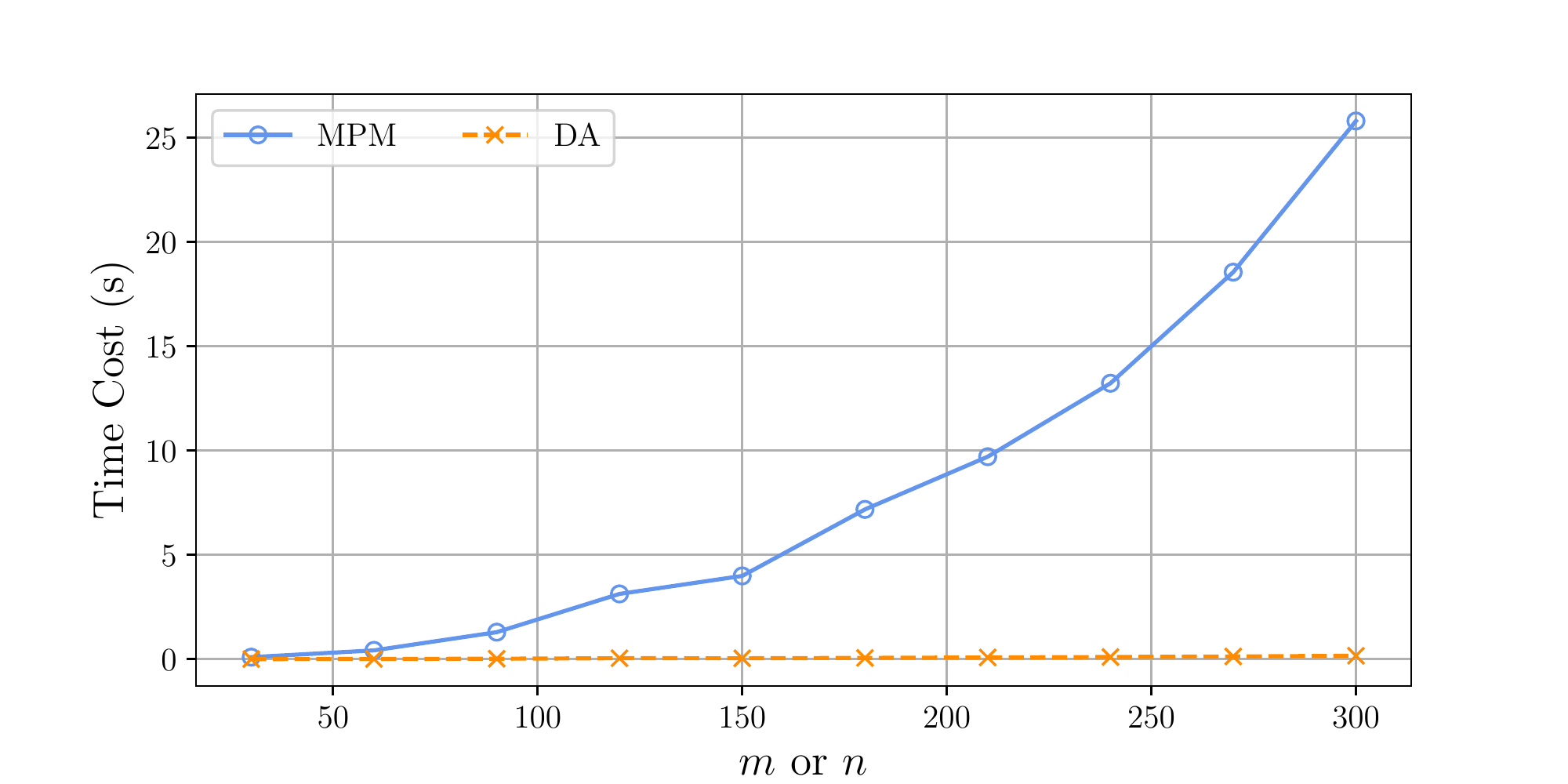}
\caption{Comparison of computational efficiency.}
\label{fig:time}
\end{figure}

Fig.~\ref{fig:time} compares the computational efficiency of MPM and DA.
We can see from the figure, the time cost of MPM rises dramatically as the number of participants increases, while DA completes within $0.2$s and its time cost barely changes.
It is because that MPM needs to solve quadratic programming problem \eqref{eq:qp}. 
Calling traditional solvers in GEKKO is the main reason for the high time consumption of MPM.
Moreover, the procedures of evaluating matching results, applying rejection rules, and generating the final matching results are also time consuming. 
On the other hand, DA does not need to solve any optimization problem or updating matching results.
Its computational cost is mainly caused by traversing requirement lists and service lists.   

\section{Conclusion}\label{sec:conclusion}

In this paper, we design a distributed computational resource trading strategy for IoT systems. The trading adopts a multi-preference matching mechanism that can well encourage the participation of collaborators with higher reputation scores.
With the help of blockchain, the decentralization of resource trading is achieved, the security, traceability, and immutability of transaction records are guaranteed, and the automation of distributed matching and reputation mechanisms is enabled.

It is worth noting that the reputation mechanism in our system is simplified for illustrative purposes.
A practical reputation system can be more comprehensive and complicated.
How to design a reasonable reputation mechanism for IoT systems in a distributed way is a direction worthy of further study. 
In addition, the evaluation of our resource trading strategy is mainly based on simulation experiments.
While experiments that considers the impact of the heterogeneity in processing time, task load, communication media, and blockchain platform in the real environment can be more helpful to test the practicability of our work.
This will take place in our future work.
Finally, computational workload prediction can help to bring more intelligence to collaborative computation task offloading.
Therefore, combining our resource trading strategy with computational workload prediction will be another direction of our future work.

\bibliographystyle{elsarticle-num-names}
\bibliography{cas-refs}



\end{document}